\documentclass[aps,prx,onecolumn,superscriptaddress,longbibliography]{revtex4-2}
\usepackage{amsmath,amsfonts,bbm}
\usepackage[colorlinks=true,breaklinks=true,linkcolor=magenta,citecolor=magenta,urlcolor=magenta]{hyperref}
\usepackage{comment}
\usepackage{graphicx}
\usepackage{floatrow}
\usepackage{mathtools}
\usepackage{booktabs}
\usepackage{diagbox}
\usepackage{listings}
\usepackage{parskip}
\usepackage{braket}
\usepackage{makecell}
\usepackage{tabularx}
\usepackage{verbatim}
\usepackage{float}
\usepackage{url}
\usepackage{algorithm}
\usepackage{algpseudocode}
\usepackage[bb=boondox]{mathalfa}
\usepackage{tikz}
\usetikzlibrary{arrows.meta}
\usetikzlibrary{quantikz}
\usetikzlibrary{calc, arrows.meta, shapes.geometric}
\usepackage{enumitem} 
\usepackage{todonotes}
\usepackage{amsmath}
\usepackage{amssymb}
\usepackage{array}
\usepackage[a4paper,top=2cm,bottom=2cm,left=2cm,right=2cm,marginparwidth=1.75cm]{geometry}
\usepackage{derivative}

\makeatletter
\def\@makecaption@fbox#1#2{%
  \noindent
  \hbox to\hsize{\hfil
    \parbox{\hsize}{%
      \noindent
      #1\@caption@fignum@sep #2%
    }%
  \hfil}%
}
\makeatother

\begin{document}

\title{Sample-based quantum simulation of vibrational structure of polyyne chains}

\author{Sarah Mostame}
\email{sarah.mostame@ibm.com}
\affiliation{IBM Research, Yorktown Heights, NY 10598, USA}

\author{Tanvi P. Gujarati}
\affiliation{IBM Research, San Jose, CA 95141, USA}

\author{Alberto Baiardi}
\affiliation{IBM Research, Rüschlikon, ZH 8803, Switzerland}

\author{Abhijit Mitra}
\affiliation{IBM Research, Austin, TX 78758, USA}

\author{Dimitar Trenev}
\affiliation{ExxonMobil Technology and Engineering Company, Annandale, NJ 08801, USA}

\author{Sumathy Raman}
\affiliation{ExxonMobil Technology and Engineering Company, Annandale, NJ 08801, USA}

\date{\today}

\begin{abstract}
Predicting anharmonic vibrational spectra is a complex computational task, which becomes quickly infeasible as molecular size and the number of vibrational basis functions increase, leading to an exponential growth of the underlying Hilbert space.
Here, we employ a sample-based quantum diagonalization (SQD) framework to compute anharmonic vibrational energies and spectra.
The approach combines quantum state preparation and sampling with configuration recovery, diagonalization of the Hamiltonian projected in a subspace of basis configurations, and calculation of transition-dipole moments to obtain both vibrational energies and infrared spectra.
We investigate two approaches for sampling, namely, vib-LUCJ and vib-SqDRIFT and apply the method to study the polyyne molecules C$_2$H$_2$ and C$_4$H$_2$, reaching representations requiring up to 104 qubits.
%For C$_4$H$_2$ with 13 vibrational modes and 8 basis functions per mode, the truncated vibrational Hilbert space has dimension $5.5\times10^{11}$, whereas the SQD calculation requires only $2.64\times10^4$ and $1.6\times10^4$ basis configurations %for vib-LUCJ ansatz and vib-SqDRIFT ansatz, respectively, 
%achieving ground-state vibrational energy difference of only $0.8~\rm{cm}^{-1}$ and $5.06~\rm{cm}^{-1}$ from the VCC[2] reference for vib-LUCJ  and vib-SqDRIFT, respectively.
Beyond energies, transition-dipole calculations reveal that infrared intensities and excited-state composition continue to evolve with basis size through anharmonic mixing and intensity redistribution, even when low-energy eigenvalues appear converged.
These results show that sample-based quantum subspace methods can capture spectroscopic information in vibrational spaces far beyond the regime of full diagonalization.
Moreover, our simulation identify multi-state recovery and scalable projected-space treatment as central challenges for further scaling sample-based approaches.
\end{abstract}

\maketitle
%\abstract {Heisenberg model}

%%%%%%%%%%%%%%%%%%%%%%%%%%%%%%%%%%%%%%%%%%%%%%%%%%%%%%%%%%%%%%%%%%%%%%%%%
%%%%%%%%%%%%%%%%%%%%%%%%%%%%%%%%%%%%%%%%%%%%%%%%%%%%%%%%%%%%%%%%%%%%%%%%%
%%%%%%%%%%%%%%%%%%%%%%%%%%%%%%%%%%%%%%%%%%%%%%%%%%%%%%%%%%%%%%%%%%%%%%%%%
\section{Introduction}
\label{sec:Introduction}

Accounting for anharmonicity in molecular simulation plays a central role in spectroscopy, thermodynamics, reaction dynamics, energy transfer, and materials characterization.
Specifically, accurate prediction of vibrational spectra requires overcoming the harmonic approximation, particularly when mode coupling, overtones, combination bands, and strongly anharmonic motions become important.
However, even neglecting the complexity of calculating highly-accurate potential energy surfaces, the computational cost of such calculations grows rapidly with both molecular size and the number of basis functions required to represent each vibrational degree of freedom. 
In product-basis approaches, the dimension of the truncated vibrational Hilbert space scales as
\begin{align}
    \dim(\mathcal{H})=\prod_{l=1}^{L} d_l \, ,
    \label{HilbertSpace_Dim}
\end{align}  
where $L$ is the number of vibrational modes and $d_l$ is the number of basis functions retained for mode $l$. 
This growth ultimately limits direct diagonalization and high-level vibrational configuration-interaction approaches for increasingly large or strongly anharmonic systems.

Several strategies have been developed to mitigate this scaling, including perturbative treatments, vibrational self-consistent field (VSCF) and configuration-interaction (CI) methods, coupled-cluster (CC) approaches, tensor decompositions, and selected-basis techniques \cite{Laidler1987_ChemicalKinetics,Sumathi2002_RateConstants,Hansen2010_VSCF,Christoffel1982_VSCFCI,Bowman1986_VSCF,Sherrill1995_CIIntroduction,Godtliebsen2013_TensorDecompositionVCC}. 
Nevertheless, obtaining accurate anharmonic vibrational states and transition properties remains a challenge when large local basis sets and multimode couplings are required. 
%The vibrational Hamiltonians relevant for practical systems, such as polyatomic hydrocarbons and reactive intermediates, often contain high-order coupling between modes and require large local basis sets for accurate description, pushing classical approaches, including perturbation theory, vibrational CI/CC methods, and tensor-network simulations, to their limits.
This motivates alternative computational strategies that utilize the structure and sparsity of physically relevant vibrational states rather than explicitly treating the complete truncated product space.
Quantum computation provides a natural representation of such finite vibrational spaces by encoding product vibrational basis states as qubit states.
By mapping bosonic degrees of freedom to qubits (e.g., via unary or binary encodings), vibrational Hamiltonians can be simulated without the need for fermion-to-qubit transformations, avoiding the long-range Jordan-Wigner tails that complicate electronic-structure simulations. 
%In contrast to electronic-structure simulations, vibrational Hamiltonians do not require a fermion-to-qubit transformation. Instead, each vibrational coordinate can be represented directly in a finite basis of one-mode functions. 
Quantum algorithms for molecular vibrations have been explored across both near-term and fault-tolerant regimes, including variational methods for vibrational eigenstates and excitation energies, quantum phase estimation for spectroscopic energy resolution, and real-time simulation of vibrational dynamics~\cite{Ollitrault2020_VibrationalVQE,Sawaya2019_VibronicSpectra,Trenev2025_TrotterResourceEstimation,Malpathak2025_BosonicDevices,Richerme2023_HydrogenBondDynamics,Lotstedt2026_RovibrationalH2O}.
A key challenge for near-term quantum implementations is that sufficiently accurate eigenstate preparation can require circuit depths beyond the capabilities of current hardware.
Sample-based quantum diagonalization (SQD) offers a complementary strategy. Rather than relying on a quantum processor to directly prepare highly accurate eigenstates, quantum-generated samples are used to identify a compact subset of physically relevant basis configurations within which the Hamiltonian is subsequently diagonalized. This shifts a portion of the computational burden from the construction of complex, coherent quantum states to the construction and solution of a substantially smaller effective eigenvalue problem~\cite{RobledoMoreno2025_SQD}.

Extending SQD to vibrational structure introduces specific challenges that are not present in electronic-structure SQD approaches: the specific symmetry structure of the vibrational Hamiltonian, the potentially large number of basis functions required to describe each mode, and the need of targeting a possibly large number of excited states (together with the corresponding transition properties) needed to calculate a full vibrational spectrum.
Notably, the calculation of transition properties, which yield spectroscopic observables, is considerably more demanding than ground-state energies alone as they depend directly on the accuracy and symmetry character of both the initial and final vibrational wavefunctions.
A selected subspace that reproduces a low-energy eigenvalue accurately may therefore still be insufficient to recover transition dipole moments and infrared intensities with a comparable accuracy.

Here, we develop an end-to-end SQD framework for optimizing the ground state of anharmonic vibrational Hamiltonians.
%represented in a truncated product basis of mode-specific vibrational functions.
The workflow combines unary encoding, two complementary sampling strategies — a vibrational adaptation of the local unitary cluster Jastrow (vib-LUCJ) ansatz and sample-based vibrational qDRIFT (vib-SqDRIFT) — with quantum sampling, configuration recovery, projected Hamiltonian construction, selected-subspace diagonalization, and transition-dipole evaluation.
We apply the approach to the polyyne systems C$_2$H$_2$ and C$_4$H$_2$, systematically increasing the number of retained basis functions per vibrational mode.
The resulting qubit representations range up to 56 qubits for C$_2$H$_2$ and 104 qubits for C$_4$H$_2$.
Our analysis addresses three complementary questions.
First, we examine whether quantum-generated samples can recover sufficiently compact subspaces to reproduce accurate low-energy vibrational eigenvalues despite the rapid growth of the complete truncated product space.
Second, we compare the behavior of vib-LUCJ and vib-SqDRIFT sampling as the local vibrational basis is enlarged.
Third, we extend the SQD workflow beyond energies to infrared transition properties and investigate whether spectroscopic observables provide a more sensitive measure of selected-subspace quality. 
For C$_4$H$_2$, the largest calculation considered here includes 13 vibrational modes with eight basis functions per mode, corresponding to a physical vibrational Hilbert-space dimension of $8^{13}\approx5.5\times10^{11}$.
Despite the size of the underlying space, vib-LUCJ and vib-SqDRIFT require selected subspaces containing only $2.64\times10^4$ and $1.60\times10^4$ product configurations, respectively, while yielding ground-state energies that remain within chemical accuracy of the VCC[2] and VCC[3] reference values~\cite{Christiansen2004_SecondQuantization,Sawaya2021_VibrationalSpectroscopyAlgorithms}, see Tables ~\ref{EnergyRef_table},  \ref{lucj_Table}, and \ref{sqdriift_Table}.

% $0.8~\mathrm{cm}^{-1}$. Vib-SqDRIFT identifies an even smaller subspace of $1.60\times10^4$ configurations, with an energy lower than the VCC[2] value by $5.06~\mathrm{cm}^{-1}$.

Beyond the energy benchmarks, infrared transition properties reveal aspects of the reconstructed vibrational wavefunctions that are not evident from eigenvalues alone.
In C$_4$H$_2$, increasing the modal basis redistributes transition strength and alters the configuration character of closely spaced excited and combination states, while the associated energies can appear comparatively well converged.
This demonstrates that spectroscopic observables provide a particularly sensitive benchmark of sample-based eigenspace reconstruction, requiring not only accurate energies but also faithful recovery of the relevant excited-state wavefunctions.

%%%%%%%%%%%%%%%%%%%%%%%%%%%%%%%%%%%%%%%%%%%%%%%%%%%%%%%%%%%%%%%%%
%%%%%%%%%%%%%%%%%%%%%%%%%%%%%%%%%%%%%%%%%%%%%%%%%%%%%%%%%%%%%%%%%
%%%%%%%%%%%%%%%%%%%%%%%%%%%%%%%%%%%%%%%%%%%%%%%%%%%%%%%%%%%%%%%%%

\section{Methods}
\label{sec:Methods}
\subsection{Hamiltonian Model}
\label{HamiltonianModel}

\begin{figure*}[t]
    \centering
    \includegraphics[width=1\linewidth]{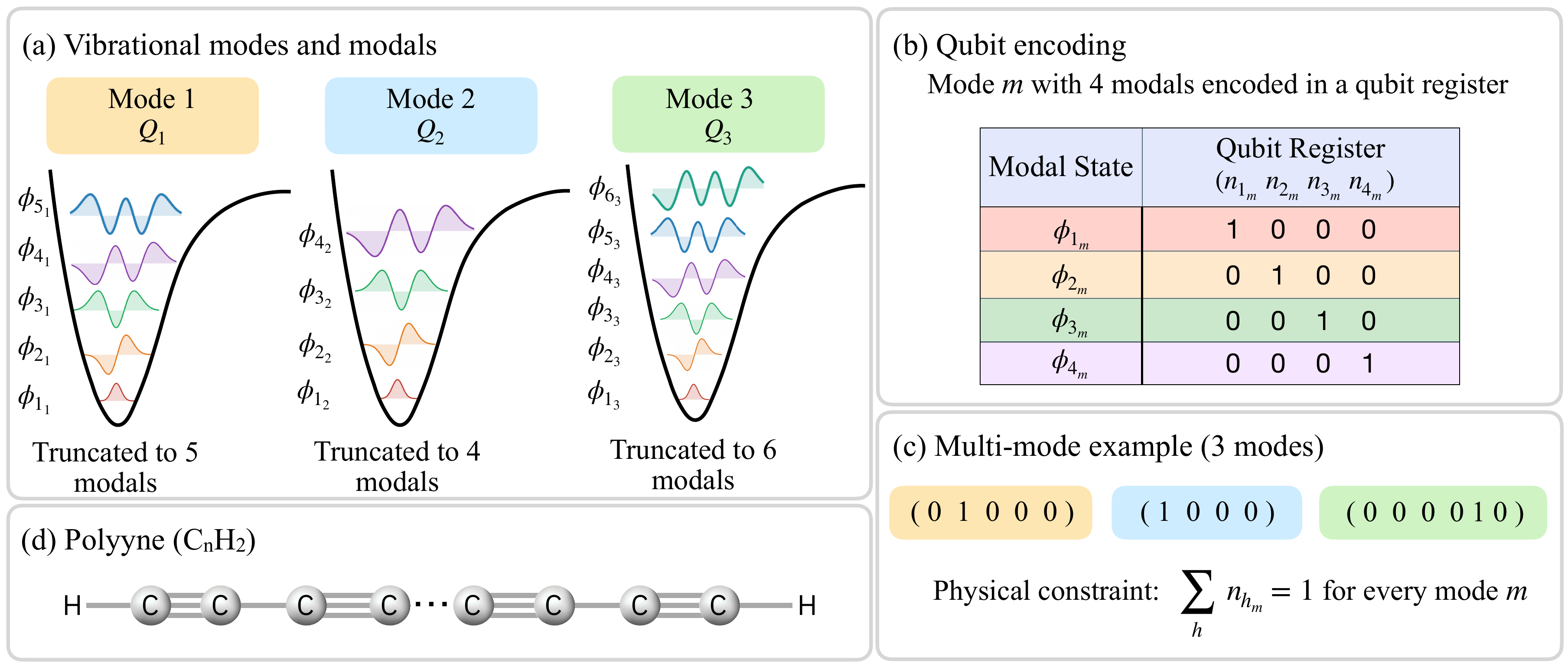}
    \caption{{\bf{Modal representation and unary qubit encoding of the vibrational Hamiltonian}}.
    (a) Each vibrational mode $Q_l$ is represented by a truncated set of mode-specific basis functions (modals), with the number of retained modals allowed to vary between modes. (b) In the unary encoding, each modal is mapped to one qubit within the register associated with a given vibrational mode. (c) A physical multi-mode configuration is formed from the tensor product of the mode registers and satisfies the constraint $\sum_h n_{h_m}=1$ for every mode $m$ corresponding to exactly one occupied modal per vibrational mode. (d) Schematic polyyne chain C$_n$H$_2$ with alternating carbon single and triple bonds. }
    \label{ModalH}
\end{figure*}

We propose optimizing the ground and excited states of the anharmonic vibrational-structure Hamiltonian on a quantum computer after mapping the vibrational creation/annihilation operators (which fulfill bosonic statistics) onto a qubit Hamiltonian via the unary mapping.
A general expression for the vibrational Hamiltonian is given by the multi-mode expansion, called $L$-mode representation, which expands the potential energy surface into a sum of one-mode potentials, two-mode potentials, three-mode potentials, and so on~\cite{Bowman2003_MULTIMODE,Li2001_HDMR,Christiansen2004_SecondQuantization}.
The Watson Hamiltonian for a system with $L$ modes of vibration each denoted by a variable $Q_{.}$ is given as:
\begin{align}
\label{eq:VibHamil}
H_{\textrm vib}(Q_1, \ldots, Q_L) &=
    - \frac{1}{2} \sum_{l=1}^{L} \frac{\partial^2}{\partial Q_l^2}
    + V(Q_1, \ldots, Q_L)
    \nonumber \\
V(Q_1, \ldots, Q_L) &=
    V_0 + \sum_{l=1}^L V^{[l]}(Q_l) 
 + \sum_{l<m}^L V^{[l,m]}(Q_l, Q_m) 
 + \sum_{l<m<r}^L V^{[l,m,r]}(Q_l, Q_m, Q_r) + \dots
\end{align}
where we have neglected the ro-vibrational contribution to the kinetic energy term.
The Hamiltonian expansion is truncated at arbitrary order $D$ which can take integer values up to $L$, the number of modes, see Figure~\ref{ModalH} for details.

The second quantization form of Eq.~\ref{eq:VibHamil} can be defined after introducing a general set of one-particle basis functions -- called modals.
To each mode, $l$, is then associated a number $d_l$ of modals $\{\phi_{n_l}\}$. 
When expressed in terms of this finite basis set, the Hamiltonian can be written as
\begin{align}
\label{eq:Hamil2ndquan}
&H_{\textrm vib}^{SQ} =
    \sum_{l=1}^L \sum_{k,h}^{d_l} \bra{\phi_{k_l}}T(Q_l) + V^{[l]}(Q_l) \ket{ \phi_{h_l} } a_{k_l}^\dagger a_{h_l} \nonumber \\    
    &+ \sum_{l<m}^L \sum_{k,h}^{d_l} \sum_{k,h}^{d_m}
    \Big\{\bra{\phi_{k_l} \phi_{k_m}}  V^{[l,m]}(Q_l, Q_m) \ket{\phi_{h_l} \phi_{h_m}} 
     \left(a_{k_l}^\dagger a_{k_m}^\dagger a_{h_l} a_{h_m} \right) \Big\} + \dots
\end{align}
where $a_{\cdot}^{\dagger}$ and $a_{\cdot}$ are the creation and annihilation operators, respectively. 
The occupation number of each bosonic mode (i.e., the number of modals per mode $d_l$) can vary, and from a chemistry perspective, the distribution of modals per mode depends on the energies of the modes. 
%
%%%%%%%%%%%%%%%%%%%%%%%%%%%%%%%%%%%%%%%%%%%%%%%%%%%%%%%%%%%%%%%%%%%%%
%%%%%%%%%%%%%%%%%%%%%%%%%%%%%%%%%%%%%%%%%%%%%%%%%%%%%%%%%%%%%%%%%%%%%
%%%%%%%%%%%%%%%%%%%%%%%%%%%%%%%%%%%%%%%%%%%%%%%%%%%%%%%%%%%%%%%%%%%%%
\subsection{Classical Pre-processing}
\label{sec:Preprocessing}
All electronic-structure calculations were performed using density functional theory with the
B3LYP exchange--correlation functional~\cite{Becke1993_ExactExchange,Lee1988_LYPCorrelation,Stephens1994_B3LYP} combined with the def2-SVP basis set~\cite{Weigend2005_Def2BasisSets}.
The harmonic force field was then obtained by numerical differentiation of the electronic energy with respect to Cartesian displacements of $10^{-3}$\,a.u.
Diagonalizing the resulting mass-weighted Hessian defined the rectilinear normal coordinates $\{Q_i\}$ used in all subsequent calculations.

The potential energy surface (PES), $V(Q)$, and the three Cartesian components
of the electric dipole moment surface, $\mu_\chi(Q)$ with
$\chi = x, y, z$, were represented by an $L$-mode expansion~\cite{Bowman2003_MULTIMODE,Kongsted2006_AutomaticForceFields} truncated at two-mode couplings.
The one- and two-mode surfaces were constructed on grids with the adaptive density-guided approach (ADGA)~\cite{Sparta2009_ADGA,Sparta2009_MultiresolutionADGA}.
ADGA iteratively extends and refines the single-point grid of each mode combination in the regions where the vibrational density is significant.
The density used to guide the ADGA is calculated through vibrational self-consistent field (VSCF).
Within the ADGA, the VSCF equations were solved on the current PES with a B-spline basis of order 20.
A mode combination was considered converged when the change in the integral of the density-weighted potential between consecutive ADGA iterations fell below a relative threshold of $10^{-2}$ and an absolute threshold of $10^{-6}\,E_\mathrm{h}$.
A maximum of 10 ADGA iterations was allowed.
With these settings, the ADGA converged for all one- and two-mode combinations of both molecules at both levels of theory.
The potential of each mode combination was fitted to a polynomial basis containing terms up to 12th order in each coordinate.
The dipole moment components were computed at the same single points as the energy and fitted in the same way.
The grids were therefore determined by the PES alone.

On the resulting surfaces, we first determined the VSCF ground state~\cite{Hansen2010_VSCF}.
For each mode, the basis boundaries were set by the classical turning points of the harmonic-oscillator state with quantum number $v = 10$.
The VSCF modals define the one-mode basis in which the Hamiltonian of Eq.~\ref{eq:Hamil2ndquan} and the dipole operators were cast in second quantization~\cite{Christiansen2004_SecondQuantization}. 
The dipole operators were transformed with the same modals as the Hamiltonian. These computations were performed using MidasCpp software package\cite{Christiansen2022_MidasCpp}.

Mode--mode correlation beyond the VSCF mean-field description was included at the
vibrational coupled-cluster level with up to two-mode excitations,
VCC[2]~\cite{Christiansen2004_VCC}, within the same modal space.
The VCC[2] equations were solved with a quasi-Newton algorithm preconditioned by the
diagonal of the VCC Jacobian.
The convergence threshold on the residual norm was $10^{-10}$\,a.u.
For the B3LYP/def2-SVP surfaces, the residual norm reached $10^{-8}$\,a.u.\ within the
iteration limit. The corresponding change in the VCC[2] energy was below
$10^{-5}$\,cm$^{-1}$, which is negligible for the present purposes.
The second-quantized Hamiltonian and dipole operators and the converged VCC[2] amplitudes were used to parametrize and initialize the vib-LUCJ circuits (see below).

\subsection{Quantum Circuit Construction}
\label{sec:QuantumCircuits}
Constructing quantum circuits for vibrational structure poses unique difficulties compared to fermionic electronic-structure problems due to the bosonic nature of vibrational excitations and the presence of higher-order couplings in the Hamiltonian. 
In this work, we investigate two ansatz strategies compatible with Sample-based Quantum Diagonalization (SQD) and suitable for near-term quantum hardware.

%%%%%%%%%%%%%%%%%%%%%%%%%%%%%%%%%%%%%%%%%%%%%%%%%%%%%%%%%%%%%%%%%%%%%%%
%%%%%%%%%%%%%%%%%%%%%%%%%%%%%%%%%%%%%%%%%%%%%%%%%%%%%%%%%%%%%%%%%%%%%%%
\subsubsection{vib-LUCJ Ansatz}
\label{sec:vibLUCJ}

We propose here a generalization of the LUCJ ansatz -- namely, vib-LUCJ -- to vibrational problems.
The key idea is to first derive the vibrational counterpart of the local unitary cluster Jastrow (LUCJ) ansatz~\cite{Motta2023_LUCJ,Matsuzawa2020_JastrowUCC} starting from the vibrational unitary coupled-cluster (VUCC) wavefunction~\cite{Ollitrault2020_VibrationalVQE} truncated to one- and two-mode excitations:
\begin{equation}
  |\Psi_\mathrm{VUCC}\rangle = e^{T-T^\dagger}|\Phi_0\rangle \, ,
  \label{eq:VUCC}
\end{equation}
where $|\Phi_0\rangle$ is the VSCF reference wave function and the excitation operator $T$ is defined as follows:
\begin{equation}
    T = \sum_{l}\sum_{h_{l},k_{l}} t_{h_{l}k_{l}}\, a^\dagger_{h_l} a_{k_l}
        + \sum_{l<m}\sum_{h_{l},k_{l}}\sum_{h_{m},k_{m}} t_{h_{l}k_{l}h_{m}k_{m}}\, a^\dagger_{h_l} a^\dagger_{h_m} a_{k_m} a_{k_l},
    \label{eq:VUCC_expanded}
\end{equation}
with $k_l$ the occupied VSCF modals of mode $l$  and $h_l$ the unoccupied modals of mode $l$.
In the unary encoding of Figure ~\ref{ModalH}, each mode carries exactly one quantum, and every operator in Eq.~(\ref{eq:VUCC_expanded}) is a product of mode-local bilinears $a^\dagger_{h_l}a_{k_l}$.
The Jordan--Wigner strings of such bilinears cancel within each mode register, so a pair of modes $(l,m)$ is exactly equivalent to a two-electron problem with a spin-$\alpha$ and a spin-$\beta$ electron, each in $d$ orbitals.
This equivalence allows the factorization machinery of the electronic LUCJ ansatz to be applied pair by pair.
Following the opposite-spin case of Refs.~\cite{Motta2021_LowRank,Motta2023_LUCJ}, the singular value decomposition of $t_{h_{l}k_{l}h_{m}k_{m}}$ writes $T_{lm}-T_{lm}^{\dagger}$ as a sum of squares of one-body operators, each diagonalized by one modal rotation per mode.
Retaining the leading term and applying a first-order Trotter splitting yields the vibrational counterpart of the LUCJ ansatz:
\begin{equation}
    |\Psi_\text{vib-LUCJ}\rangle %= e^{T_1-T_1^\dagger}
        = \prod_{(l,m)\in\mathcal{P}} e^{K_{lm}}\, e^{iJ_{lm}}\, e^{-K_{lm}}\, |\Phi_0\rangle, 
        \qquad \text{with} \qquad
        J_{lm} = \sum_{k,h} J^{lm}_{kh}\, n_{k_l} n_{h_m},
    \label{eq:vibLUCJ}
\end{equation}
where $e^{K_{lm}}$ is a product of two one-mode modal rotations acting on modes $l$ and $m$, and $n_{k_l}=a^\dagger_{k_l}a_{k_l}$.
Since VCC[2] contains only one- and two-mode excitations, $J_{lm}$ is purely two-mode.
Following the approach used in electronic-structure calculations~\cite{RobledoMoreno2025_SQD}, we select the parameters for the vib-LUCJ ansatz by executing first a VCC[2] calculation and converting the converged VCC[2] amplitudes (Sec.~\ref{sec:Preprocessing}) to the entries of the $K_{lm}$ and $J_{lm}$ operators.
Notably, the vib-LUCJ circuit never leaves the physical subspace.
Measured bitstrings that violate this constraint are therefore due to hardware noise, and can be corrected by configuration recovery~\cite{RobledoMoreno2025_SQD}.

%ALB This can be uncommented once we actually do that.
%Because the pair unitaries do not commute, the product in Eq.~(\ref{eq:vibLUCJ}) is ordered.
%Pairs with the same separation $|l-m|$ that share no mode are grouped into parallel layers, which are applied in order of increasing separation.
%Before building the circuit, the modes are relabeled according to the Fiedler ordering~\cite{Fiedler1973_AlgebraicConnectivity,Barcza2011_OrbitalOrdering} of the mode-coupling graph with weights $w_{lm}=\sum_{ab}|t^{lm}_{ab}|$, which places strongly coupled modes next to each other in the qubit register.
%In the electronic LUCJ ansatz, the locality restriction retains only Jastrow couplings between orbitals that are adjacent on the hardware~\cite{Motta2023_LUCJ}.
%Here it is imposed in mode space instead: $\mathcal{P}$ contains only the pairs with $|l-m|\le k$ in Fiedler order.
%For $k=1$, the number of pairs drops from $L(L-1)/2$ to $L-1$, i.e., from 21 to 6 for C$_2$H$_2$ and from 78 to 12 for C$_4$H$_2$.
The prepared state generates the sampling distribution $p(\mathbf{n})=|\langle\mathbf{n}|\Psi_\text{vib-LUCJ}\rangle|^2$ over modal product configurations $|\mathbf{n}\rangle$.
In the spirit of SQD-based approaches, the sampled configurations define the subspace in which the Hamiltonian is projected and diagonalized.
The vib-LUCJ combined with SQD therefore realizes the quantum analogue of selected vibrational configuration interaction, in which configurations are sampled from a quantum circuit rather than by perturbative criteria.

%%%%%%%%%%%%%%%%%%%%%%%%%%%%%%%%%%%%%%%%%%%%%%%%%%%%%%%%%%%%%%%%%%%%%
%%%%%%%%%%%%%%%%%%%%%%%%%%%%%%%%%%%%%%%%%%%%%%%%%%%%%%%%%%%%%%%%%%%%%
\subsubsection{vib-SqDRIFT Ansatz}
\label{sec:vibSqDRIFT}
An alternative to the LUCJ ansatz to obtain the support of bit-strings for diagonalization is to sample from quantum circuits generated by approximating the time evolution of the Hamiltonian. In particular, sample-based Krylov diagonalization \cite{Yu2025_SampleBasedKrylov} and sample-based qDRIFT diagonalization \cite{Piccinelli2026_SqDRIFT}, a.k.a. SqDRIFT have been explored in the space of fermionic Hamiltonians. These methods are derived from the prior work on Krylov subspaces \cite{Parrish2019_QuantumFilterDiagonalization} and the qDRIFT algorithm developed by Campbell \cite{Campbell2019_qDRIFT}. 
%{ToDo: Sentence on convergence guarantees} It was shown in \cite{Piccinelli2026_SqDRIFT} that the convergence guarantees similar to those of phase estimation are preserved for SqDRIFT as long as the assumptions of good overlap with the ground state and polynomially growing support of the desired eigenstates with system size are satisfied.

We extend the workflow developed for SqDRIFT to simulate time evolution of the vibrational Hamiltonian. For the given vibrational Hamiltonian in Eq. (\ref{eq:Hamil2ndquan}) represented as, $H = \sum_j c_jH_j$, we sample $N_{exc}$ terms from a probability distribution proportional to operator norms $\|c_jH_j\|$ with replacement. A given initial state is then time evolved with the sampled Hamiltonian operator terms for a common time step $\delta=\lambda t/N_{exc}$, where $\lambda =\sum_j \|c_jH_j\|$ is the total spectral weight and $t$ is the target evolution time. A noteworthy feature of the qDRIFT approach is that the error incurred in approximating the exact time evolution with this method is independent of the total number of terms in the Hamiltonian and is dependent on the square of the one norm, $\lambda$. 

The randomly sampled time evolution operator after mapping to the qubit representation is simulated on a quantum computer and modal configurations are measured in the one-particle modal basis for $R$ realizations of randomly sampled $N$ Hamiltonian terms. This is repeated for different Krylov factors, $k$, targeting total evolution times of \{$k*t$\} where $t$ defines the base evolution time. Measured modal configurations for a fixed value of $N_{exc}$ obtained from different randomizations for all values of $k$ are combined together to generate pool of configurations used in the sample-based diagonalization. An in-depth discussion of the SqDRIFT procedure and corresponding convergence guarantees can be found in \cite{Piccinelli2026_SqDRIFT}. Similar to the grouping of terms in a Hamiltonian that preserve symmetry properties discussed in \cite{Piccinelli2026_SqDRIFT}, we perform a grouping of the Bosonic terms of the vibrational Hamiltonian such that each Bosonic term is paired with its Hermitian adjoint and treated as a single block for sampling of the Hamiltonian terms. Any small numerical asymmetries between adjoint coefficients are removed by averaging them. Every block is assigned a sampling weight equal to its coefficient one-norm, $w_j=\sum_a \|c_{ja}\|$, and each block is normalized by dividing through by $w_j$ before being mapped to qubit space using the direct mapping~\cite{Ollitrault2020_VibrationalVQE}, which encodes each modal directly as a qubit. %The normalized Pauli groups are stored once, and multiple independent SqDRIFT randomizations are generated by drawing $N$ block indices with replacement from the probability distribution $p_j=w_j/\lambda$, each randomization using a deterministic seed derived from a common master seed to ensure full reproducibility. For each randomization, the ordered sequence of normalized SparsePauliOp blocks is assembled into a Qiskit quantum circuit in which each sampled block contributes one PauliEvolutionGate applied for the shared time step $\delta =\lambda t/N$. 
After mapping to qubits, approximated time evolution circuits were constructed based on Lie-Trotter product formula with a single time step using Qiskit \cite{JavadiAbhari2024_Qiskit} native functions. The generated circuits were then executed on the \texttt{ibm\_phoenix} device featuring NighthawkR2 quantum processor. For both the systems, $\text{C}_{2}\text{H}_{2}$ and $\text{C}_{4}\text{H}_{2}$, circuits for $N_{exc}$ = $\{25,50,100,250\}$ excitations (operator terms) each with $R$ = 100 randomizations were constructed. Each circuit for a fixed $N_{exc}$ excitations was run with five Krylov time steps of $k$ = $\{1,2,3,4\}$ and a base time $t$ given by $\min(1.0, \pi/\lambda)$. The initial VSCF state corresponding to the $k$ = 0 vector was added by default in the subspace of configurations before diagonalization. Each circuit was sampled on the quantum hardware with 5000 shots, thus leading to $2.0 \times 10^6$ measurement shots for every $N_{exc}$. Dynamical Decoupling with the XpXm sequence provided via Qiskit was added for error suppression to every circuit. The average required circuit resources for all the randomizations are reported in Appendix \ref{Circuit_Resources}

\subsection{Classical Post-Processing}
\label{sec:ClassPostProcessing}

Even when the applied quantum unitary operators of the ansatz are structured to preserve a single modal occupancy per mode, due to the noise present in the quantum hardware, the obtained samples can be corrupted. To tackle the effect of noise, we apply the iterative configuration recovery protocol introduced in \cite{RobledoMoreno2025_SQD} along with the carryover functionality introduced in \cite{shirakawa2025closedloop}, but adapted for estimating ground and excited states with restrictions imposing single modal occupancy. For obtained HW counts the 0th iteration of the configuration recovery filters the counts with the right modal occupancies and creates $B$ independent batches with $n_{B}$ samples per batch. The given Hamiltonian is then projected in the subspace generate by the sampled configurations and diagonalized for each batch independently. Classical diagonalization subroutines from scipy were used for this step and the ground and excited states were computed. Modal occupations generated from the lowest energy ground state among the batches is used to seed the input occupancy for the next configuration recovery iteration. %ToDo: Details about the configuration recovery subroutine. 
Independent batches are then sampled again from the corrected counts and the diagonalization process is repeated. In addition to sampling from the corrected bitstrings, bitstrings present in the ground and excited states of the previous iteration with weight greater than the carryover threshold of $1e^{-4}$ are added as default in the subspace for projection of the Hamiltonian. The configuration recovery loop along with bitstring carryover is repeated until convergence in ground state energies and occupancies is reached or a fixed number of iterations are exhausted. The final lowest ground state energies are reported and ground and excited state wavefunctions are used to compute the transition dipole moments and infrared spectra as discussed in the next section.

%%%%%%%%%%%%%%%%%%%%%%%%%%%%%%%%%%%%%%%%%%%%%%%%%%%%%%%%%%%%%%%%%%%%%
\subsection{Intensity and transition dipole moment}
\label{sec:transition dipole}

In highly symmetric molecules, especially quasi-1D systems like polyynes, the vibrational intensities contain more symmetry information than the energies. The frequencies may look nearly harmonic and featureless, while the transition intensities reveal hidden collective symmetry, localization, inversion breaking, Peierls distortions, vibronic coupling, and low-frequency modes~\cite{NISTWebBook,chalifoux2010synthesis,Agarwal2013,milani2008}. 
Ideal polyynes are linear carbon chains HC$_{2n}$H with approximate symmetry  $D_{\infty h}$. They have inversion symmetry, quasi-1D translational symmetry, nearly degenerate collective stretching modes, strong electron-phonon coupling, and long-range vibrational correlations.  
The low-frequency vibrations are often collective longitudinal modes, chain-bending modes, breathing-like distortions, Peierls-like dimerization fluctuations, and symmetry-breaking fluctuations.  
These modes may be nearly invisible in energies alone. 
Diagonalizing the vibrational Hamiltonian $H_{\textrm vib} \ket{\Psi_l} = E_l \ket{\Psi_l}$, we can obtain the energy spectrum $E_n$ while the intensity probe $\bra{\Psi_m} \mu \ket{\Psi_l}$ or $\bra{\Psi_m} \alpha \ket{\Psi_l}$ could encode parity, localization, phase structure, mode mixing and hidden symmetry sectors with 
$\mu = - {\partial E}/{\partial F}$ being the dipole moment (the first derivative of energy) and 
$\alpha = - {\partial^2 E}/{\partial F^2}$ being the polarizability (the second derivative of energy), while $F$ is the external electric field.

Each vibrational mode $l$ has a modal basis $\{\phi_{k_l}^l (Q_l) \}$.
Full vibrational basis is a product basis 
\begin{align}
\label{eq:VibBase}
\ket{\mathcal{K}} =\ket{k_1 k_2 \cdot k_L} =  \prod_{l=1}^L \phi_{k_l}^l (Q_l)\, , 
\end{align}
where $k_l$ labels the modal of that mode occupied. 
After diagonalizing the Hamiltonian $H_{\textrm vib} \ket{\Psi_l} = E_l \ket{\Psi_l} $, each eigenstate would be  $\ket{\Psi_l} = \sum_{\mathcal K} C_{\mathcal K}^l \, \ket{\mathcal K}$ with frequencies given as $\omega_{ml} = (E_m - E_l)/\hbar\, .$

To calculate the intensities, similar to the potential energy surface in Eq.~\ref{eq:VibHamil}, the dipole moment surface is represented with 
\begin{align}
\label{eq:dipolemoment}
\mu_\chi(Q_1, \ldots, Q_L) =
    \mu_{\chi,0} + \sum_{l=1}^L \mu_\chi^{[l]}(Q_l) 
 + \sum_{l<m}^L \mu_\chi^{[l,m]}(Q_l, Q_m) 
 + \sum_{l<m<r}^L \mu_\chi^{[l,m,r]}(Q_l, Q_m, Q_r) + \cdots
\end{align}
with $\chi = x, y, z $, which identifies the molecular dipole moment when the nuclei are displaced to a particular geometry.
In the second quantization, the same way as the vibrational Hamiltonian in Eq.~\ref{eq:Hamil2ndquan}, this becomes
\begin{align}
\label{eq:dipole2ndquan}
\mu_\chi = \mu_{\chi,0} + 
             \sum_{l=1}^L \sum_{k,h}^{d_l} \mu_{k_l , h_l}^{\chi,l} a_{k_l}^\dagger a_{h_l} &+ \sum_{l<m}^L \sum_{k,h}^{d_l} \sum_{k,h}^{d_m} \mu_{k_l , h_l, k_m , h_m}^{\chi,l,m} a_{k_l}^\dagger a_{k_m}^\dagger a_{h_l} a_{h_m} 
             + \cdots \, , 
\end{align}
where $a_{\cdot}^{\dagger}$ and $a_{\cdot}$ are the creation and annihilation operators, respectively, and 
\begin{align}
\label{eq:dipole_modal_integral}
    \mu_{k_l , h_l}^{\chi,l} = \int dQ_l \, \phi_{k_l}^l(Q_l)\, \,  \mu_{\chi,l} (Q_l) \,  \phi_{h_l}^l(Q_l), 
    \quad 
    \mu_{k_l , h_l, k_m , h_m}^{\chi,l,m} = \int dQ_l \, dQ_l \,  \phi_{k_l}^l\, \phi_{k_m}^m\, \,  \mu_{\chi,l, m} (Q_l, Q_m)\, \phi_{h_l}^l \, \phi_{h_m}^m
\end{align}
are the modal integrals. 
The transition dipole moment from vibrational state $l$ to $m$ is  given by
\mbox{$\mu_\chi^{ml}=  \bra{\Psi_m}\mu_\chi\ket{\Psi_l} $,}
component-wise for $\chi = x, y, z $. Using modal coefficients
\begin{align}
    \label{eq:transitionDipole}
    \mu_\chi^{ml}=   \sum_{\mathcal K, \mathcal J} \left(C_{\mathcal K}^m\right)^* C_{\mathcal J}^l\, \bra{\mathcal K}\mu_\chi\ket{\mathcal J} 
\end{align}
Where $\mathcal K$ and  $\mathcal J$ are the vibrational basis given in Eq.~\ref{eq:VibBase}. 
Then the total transition dipole strength is 
\begin{align}
    \label{eq:dipoleStrength}
    {\mathcal S}_{ml}=   \sum_{\chi = x, y, z  } \left|\mu_\chi^{ml}\right|^2\, .
\end{align}

Given a $l$-mode vibrational Hamiltonian in a modal basis, the infrared (IR) intensities are obtained by constructing a $l$-mode expansion of the dipole moment surface in the same modal basis and evaluating transition dipole matrix elements between the computed vibrational eigenstates. 
Therefore the IR absorption intensity is proportional to 
\begin{align}
    \label{eq:absorptionIntensity}
    I_{l\rightarrow{m}}^{\textrm{IR}} \propto \, \omega_{ml} \sum_{\chi = x, y, z  } \left|\bra{\Psi_m}\mu_\chi\ket{\Psi_l} \right|^2   
\end{align}
where at zero temperature, usually $l=0$, and then 
$I_{0\rightarrow{m}}^{\textrm{IR}} \propto \, \omega_{m0} \sum_{\chi} \left|\bra{\Psi_m}\mu_\chi\ket{\Psi_0} \right|^2 $. 
At finite temperature, 
\begin{align}
    \label{eq:finiteT_absorptionIntensity}
    I^{\textrm{IR}} (\omega) \propto \,\sum_{l,m}\, P_l \,  \omega_{ml} \sum_{\chi } \left|\bra{\Psi_m}\mu_\chi\ket{\Psi_l} \right|^2  \, \delta(\omega - \omega_{ml}) 
\end{align}
where $P_l = e^{-\beta E_l}/Z$, with $Z= \sum_l e^{-\beta E_l}$ being the partition function, $\beta = 1/(k_B T)$ where $k_B$ is the Boltzmann's constant and $T$ is the temperature. 
%

%The Raman intensity follows the same steps, except that we would need to replace the dipole moment surface by the polarizability surface $\alpha_{\chi\gamma}$
%\begin{align}
%    \label{eq:RamanIntensity}
%    I_{0\rightarrow{m}}^{\textrm{Raman}} \propto \,
%    \left|\bra{\Psi_m}\alpha_{\chi\gamma}\ket{\Psi_0} \right|^2  
%\end{align}
%where $\chi,\gamma = x, y, z $ are the Cartesian directions that label the polarizability tensor. 

%%%%%%%%%%%%%%%%%%%%%%%%%%%%%%%%%%%%%%%%%%%%%%%%%%%%%%%%%%%%%%%%%%%%%%%%%
%%%%%%%%%%%%%%%%%%%%%%%%%%%%%%%%%%%%%%%%%%%%%%%%%%%%%%%%%%%%%%%%%%%%%%%%%
%%%%%%%%%%%%%%%%%%%%%%%%%%%%%%%%%%%%%%%%%%%%%%%%%%%%%%%%%%%%%%%%%%%%%%%%%
%%%%%%%%%%%%%%%%%%%%%%%%%%%%%%%%%%%%%%%%%%%%%%%%%%%%%%%%%%%%%%%%%%%%%%%%%
%%%%%%%%%%%%%%%%%%%%%%%%%%%%%%%%%%%%%%%%%%%%%%%%%%%%%%%%%%%%%%%%%%%%%%%%%
%%%%%%%%%%%%%%%%%%%%%%%%%%%%%%%%%%%%%%%%%%%%%%%%%%%%%%%%%%%%%%%%%%%%%%%%%
\section{Results}

\subsection{Validation of the vibrational SQD workflow}
\label{SQD_Validation}
The first objective is to validate the complete computational workflow connecting the modal vibrational Hamiltonian, quantum sampling, configuration recovery, and projected Hamiltonian diagonalization. Particular attention was paid to the modal ordering because the circuit construction employs a Fiedler reordering of the vibrational modes to reduce circuit complexity. Before configuration recovery, measured bitstrings were transformed back to the original MIDAS ordering~\cite{Christiansen2022_MidasCpp}. Round-trip permutation tests verified that the coded transformation is exactly invertible, and independent reconstruction of the Hamiltonian on the recovered configuration basis confirmed the consistency between the sampled configurations and the projected Hamiltonian.

For modal spaces in which direct diagonalization remains feasible, we additionally compare the SQD results against results obtained from exact diagonalization of the same truncated Hamiltonian. 
This comparison is important because it separates errors associated with sample-based subspace reconstruction from those arising from the finite modal basis itself. 
The agreement between the SQD and the exact spectra shown in Figures ~\ref{DipoleC2H2_lucj}-\ref{Dipole_C4H2_SqDrift} for the classically accessible cases provides an independent validation of the recovered eigenspace and of the subsequent transition-dipole calculation. 
In particular, the close agreement of both transition positions and dominant infrared features demonstrates that the sampled subspace retains not only the low-energy eigenvalues but also the wavefunction information required to evaluate spectroscopic observables.

%%%%%%%%%%%%%%%%%%%%%%%%%%%%%%%%%%%%%%%%%%%%%%%%%%%%%%%%%%%%%%%%%%%%%%%%%
%%%%%%%%%%%%%%%%%%%%%%%%%%%%%%%%%%%%%%%%%%%%%%%%%%%%%%%%%%%%%%%%%%%%%%%%%
%%%%%%%%%%%%%%%%%%%%%%%%%%%%%%%%%%%%%%%%%%%%%%%%%%%%%%%%%%%%%%%%%%%%%%%%%
%%%%%%%%%%%%%%%%%%%%%%%%%%%%%%%%%%%%%%%%%%%%%%%%%%%%%%%%%%%%%%%%%%%%%%%%%

\subsection{Energy convergence and selected-subspace compression}
\label{EnergyConvergence}
The key advantage of the sample-based approach becomes apparent as the number of basis functions per vibrational mode is increased. The dimension of the complete physical product space grows as $d^L$, whereas the number of configurations retained after sampling and configuration recovery remains several orders of magnitude smaller. This separation is particularly noticeable for C$_4$H$_2$, which contains 13 vibrational modes. Increasing the local basis from two to eight modals per mode enlarges the physical Hilbert-space dimension from $2^{13}=8192$ to $8^{13}=5.50\times10^{11}$. In contrast, the corresponding vib-LUCJ selected spaces contain only 118 and 26437 product configurations, respectively. At eight modals per mode, the vib-LUCJ subspace therefore represents approximately $4.8\times10^{-8} $ of the complete physical product space, corresponding to a reduction by approximately $2.1\times10^7$. The vib-SqDRIFT approach achieves a comparable reduction in the number of product configurations. 
Despite this substantial compression, the deviations of the resulting ground-state vibrational energies from the VCC[2] and VCC[3] reference values remain within the 1–10~$\mathrm{cm}^{-1}$ accuracy range considered relevant for vibrational spectroscopy~\cite{Sawaya2021_VibrationalSpectroscopyAlgorithms}.
%is $8215.96~\mathrm{cm}^{-1}$, only $0.80~\mathrm{cm}^{-1}$ below the eight-modal VCC[2] reference of $8216.76~\mathrm{cm}^{-1}$. 
Tables~\ref{lucj_Table} and \ref{sqdriift_Table} summarize the vib-LUCJ and vib-SqDRIFT results for C$_2$H$_2$ and C$_4$H$_2$ with 2, 4, 6, and 8 modals per mode, together with comparisons to the VCC[2] and VCC[3] reference energies listed in Table~\ref{EnergyRef_table}. 

%A similar compression is obtained with vib-SqDRIFT. For the eight-modal C$_4$H$_2$ calculation, only 16039 configurations are retained from the $5.50\times10^{11}$-dimensional physical space, corresponding to a fraction of approximately $2.9\times10^{-8}$, or a reduction by $3.4\times10^7$. The resulting ground-state energy is $8211.70~\mathrm{cm}^{-1}$, differing from the eight-modal VCC[2] reference by $5.06~\mathrm{cm}^{-1}$.

The rapid decrease in the fraction of the truncated physical product space retained by SQD as the modal basis is enlarged is consistent with the low-energy states being concentrated in a highly structured subset of the full basis. The absolute selected-space dimension, however, does not increase monotonically with the number of modals $d$. This behavior should not be interpreted as a reduction in the intrinsic complexity of the vibrational problem, since the recovered subspace also depends on the sampling distribution, sampling budget, configuration-recovery parameters, and convergence criteria. The important observation is that low-energy vibrational energies can be recovered with selected spaces that constitute only a very small fraction of the full truncated product basis, without explicitly constructing or diagonalizing the complete space.

%%%%%%%%%%%%%%%%%%%%%%%%%%%%%%%%%%%%%%%%%%%%%%%%%%%%%%%%%%%%%%%%%%%%%%%%%
%%%%%%%%%%%%%%%%%%%%%%%%%%%%%%%%%%%%%%%%%%%%%%%%%%%%%%%%%%%%%%%%%%%%%%%%%
\begin{table}[t]
\centering
\caption{
\label{EnergyRef_table}
Reference VCC[2] and VCC[3] vibrational energies for C$_2$H$_2$ and C$_4$H$_2$ calculated using 2, 4, 6, and 8 modals per vibrational mode.  }
\begin{tabular}{|c|c|c|c|c|} 
    \hline
    Molecule & modes & modals & VCC[2]~cm$^{-1}$  & VCC[3]~cm$^{-1}$  \\ \hline\hline
C$_2$H$_2$ & 7 & 2 & 5940.04 & 5940.04   \\  \hline 
C$_2$H$_2$ & 7 & 4 & 5885.75 & 5884.47 \\ \hline  
C$_2$H$_2$ & 7 & 6 & 5883.95 & 5881.89 \\ \hline
C$_2$H$_2$ & 7  & 8 & 5883.85 & 5881.50 \\ \hline
C$_4$H$_2$ & 13 & 2 & 8297.68 & 8297.53  \\ \hline 
C$_4$H$_2$ & 13 & 4 & 8220.58 & 8216.82\\ \hline  
C$_4$H$_2$ & 13 & 6 & 8217.03 & 8210.42  \\ \hline  
C$_4$H$_2$ & 13 & 8 & 8216.76 & 8208.25 \\ \hline  
  \end{tabular}
\end{table}

%%%%%%%%%%%%%%%%%%%%%%%%%%%%%%%%%%%%%%%%%%%%%%%%%%%%%%%%%%%%%%%%%%%%%%%%%
\begin{table}[t]
\centering
\caption{
\label{lucj_Table}
\textbf{SQD results with the vib-LUCJ ansatz:} Number of qubits for the unary encoding $N_q = Ld$, with $L$ being  number of the vibrational modes and $d$ number of the modals. Number of the vibrational modes is given by $L=3N-6$ for nonlinear molecules and $L=3N-5$ for the linear ones, with $N$ being the number of atoms. The SQD results were based on 5 batches, 5000 samples per batch,  100 eigen-states and up to 20 configuration recovery iterations (if needed) for all the simulation except for C$_4$H$_2$  with 8 modals. For the latter the configuration recovery was run with 20000 samples per batch, 50 eigen-states and 50 iterations to obtain the results. All the energies are in $\mathrm{cm}^{-1}$ and SQD energies are compared to VCC[2] and VCC[3] energies. SSDim denotes the number of product-basis configurations retained in the selected subspace.}
\begin{tabular}{|c|c|c|c|c|c|c|c|c|} 
    \hline
    Molecule & modes & modals & qubits   & $E_{\rm SQD} $ & $E_{\rm SQD}$  - $E_{\rm VCC2}$ & $E_{\rm SQD}$  - $E_{\rm VCC3}$ & SSDim & Max.~SSDim \\ \hline\hline
C$_2$H$_2$ & 7 & 2 & 14  & 5940.04 & 0 & 0 & 10 & 128 \\  \hline 
C$_2$H$_2$ & 7 & 4 & 28  & 5884.46  & -1.29 & -0.01 & 2381 & 16384 \\ \hline  
C$_2$H$_2$ & 7 & 6 & 42  & 5881.83  & -2.12 & -0.06 & 8861 & 279936 \\ \hline
C$_2$H$_2$ & 7  & 8 & 56 & 5881.54 & -2.31 & 0.04 & 12216 & 2097152 \\ \hline
C$_4$H$_2$ & 13 & 2 & 26 & 8299.50 & 1.82 & 1.97 & 118 & 8192 \\ \hline 
C$_4$H$_2$ & 13 & 4 & 52 & 8217.56 & -3.02 & 0.74 & 18177 & 67108864\\ \hline  
C$_4$H$_2$ & 13 & 6 & 78 & 8216.21 & -0.82 & 5.79 & 12216 &  1.306E+10 \\ \hline  
C$_4$H$_2$ & 13 & 8 & 104 & 8215.96 & -0.80 & 7.71 & 26437 &  5.498E+11  \\ \hline  
  \end{tabular}
\end{table}
%%%%%%%%%%%%%%%%%%%%%%%%%%%%%%%%%%%%%%%%%%%%%%%%%%%%%%%%%%%%%%%%%%%%%%%%%
%%%%%%%%%%%%%%%%%%%%%%%%%%%%%%%%%%%%%%%%%%%%%%%%%%%%%%%%%%%%%%%%%%%%%%%%%
\begin{table}[h]
    \centering
  \begin{tabular}{|c|c|c|c|c|c|c|c|c|c|} 
    \hline
    Molecule & modes & modals & qubits  &  $N_{exc}$ & $E_{\rm SQD}$ & $E_{\rm SQD}$  - $E_{\rm VCC2}$ & $E_{\rm SQD}$  - $E_{\rm VCC3}$ & SSDim & Max. SSDim\\ \hline\hline  
C$_2$H$_2$ & 7 & 2 & 14 & 25 & 5940.04 & 0 & 0 & 93 & 128 \\ \hline 
C$_2$H$_2$ & 7 & 4 & 28 & 250 & 5884.46 & -1.29 & -0.01 & 6345 & 16384 \\ \hline  
C$_2$H$_2$ & 7 & 6 & 42 & 50 & 5881.82 & -2.13 & -0.07 & 11688 & 279936 \\ \hline
C$_2$H$_2$ & 7  & 8 & 56 & 250 & 5881.38 & -2.47 & -0.12 & 20591 & 2097152\\ \hline
C$_4$H$_2$ & 13 & 2 & 26 & 250 & 8297.53  & -0.15 & 0 & 5035 & 8192 \\ \hline  
C$_4$H$_2$ & 13 & 4 & 52 & 50 & 8216.85 & -3.73 & 0.03 & 23734 & 67108864 \\ \hline  
C$_4$H$_2$ & 13 & 6 & 78 & 25 & 8211.86 & -5.17 & 1.44 & 18729 & 1.306E+10\\ \hline   
C$_4$H$_2$ & 13 & 8 & 104 & 25 & 8211.70 & -5.06 & 3.45  & 16039 & 5.498E+11  \\ \hline  
  \end{tabular}
\caption{\textbf{SQD results with the vib-SqDRIFT ansatz:} Number of qubits for the unary encoding $N_q = Ld$, with $L$ being  number of the vibrational modes and $d$ number of the modals. Number of the vibrational modes is given by $L=3N-6$ for nonlinear molecules and $L=3N-5$ for the linear ones, with $N$ being the number of atoms. 
Lowest energies obtained out of all excitations $N_{exc}$ are reported along with the corresponding $N_{exc}$ value. The SQD results were based on 5 batches, 5000 samples per batch, 100 eigen-states states and up to 20 configuration recovery iterations for the C$_2$H$_2$ simulations with 2,4 and 6 modals; for 8 modals, 7500 samples per batch and 30 configuration recovery iterations were used instead. For C$_4$H$_2$ with 2 modals SQD results were based on 5 batches, 5000 samples per batch, 50 eigen-states states and up to 20 configuration recovery iterations. For C$_4$H$_2$ with 4,6 and 8 modals, SQD results were based on 5 batches, 7500 samples per batch, 50 eigen-states and up to 30 configuration recovery iterations. All the energies are in $\mathrm{cm}^{-1}$ and SQD energies are compared to VCC[2] and VCC[3] energies. SSDim denotes the number of product-basis configurations retained in the selected subspace.}
\label{sqdriift_Table}
\end{table}

%%%%%%%%%%%%%%%%%%%%%%%%%%%%%%%%%%%%%%%%%%%%%%%%%%%%%%%%%%%%%%%%%%%%%%%%%
%%%%%%%%%%%%%%%%%%%%%%%%%%%%%%%%%%%%%%%%%%%%%%%%%%%%%%%%%%%%%%%%%%%%%%%%%
\subsection{Comparison of vib-LUCJ and vib-SqDRIFT}
\label{comparison}
Both sampling strategies generate selected spaces that remain substantially smaller than the complete vibrational Hilbert space, although their behavior differs as the modal basis is enlarged. For C$_2$H$_2$, vib-LUCJ and vib-SqDRIFT yield comparable ground-state energies across the investigated basis sizes. For C$_4$H$_2$, both approaches likewise recover compact selected subspaces relative to the rapidly growing full product space. However, the differences between the two methods become more pronounced at larger modal truncations.
For the six- and eight-modal cases, the relative performance of the two sampling strategies depends on the reference method: vib-SqDRIFT exhibits larger deviations from VCC[2], while vib-LUCJ exhibits larger deviations from VCC[3].

These differences should not be interpreted as a direct ranking of the two state-preparation strategies because the sampling protocols are not matched in quantum or classical cost. In particular, the number of SqDRIFT excitations, samples per batch, and configuration-recovery iterations varies across the calculations. Rather, the results demonstrate that qualitatively different sampling mechanisms can identify compact vibrational subspaces capable of recovering low-energy physics. A controlled comparison at fixed sampling cost will be necessary to determine their relative efficiency. 
For some modal truncations, the SQD ground-state energies lie below the corresponding VCC[2] and VCC[3] values, see Tabels~\ref{lucj_Table} and \ref{sqdriift_Table}. Since SQD obtains the lowest eigenvalue of the Hamiltonian projected onto the selected subspace, it is variational within that subspace, whereas VCC[2] and VCC[3] are approximate, non-variational methods. Consequently, a lower SQD energy relative to VCC[2] and VCC[3] while encouraging, should be interpreted with caution across multiple systems. Where computationally feasible, comparison with more systematic classical benchmarks, such as VCI within the same modal basis, would provide a more definitive assessment of the SQD energies.

%%%%%%%%%%%%%%%%%%%%%%%%%%%%%%%%%%%%%%%%%%%%%%%%%%%%%%%%%%%%%%%%%%%%%%%%%
%%%%%%%%%%%%%%%%%%%%%%%%%%%%%%%%%%%%%%%%%%%%%%%%%%%%%%%%%%%%%%%%%%%%%%%%%
\begin{figure*}[h]
    \centering
    \includegraphics[width=1\linewidth]{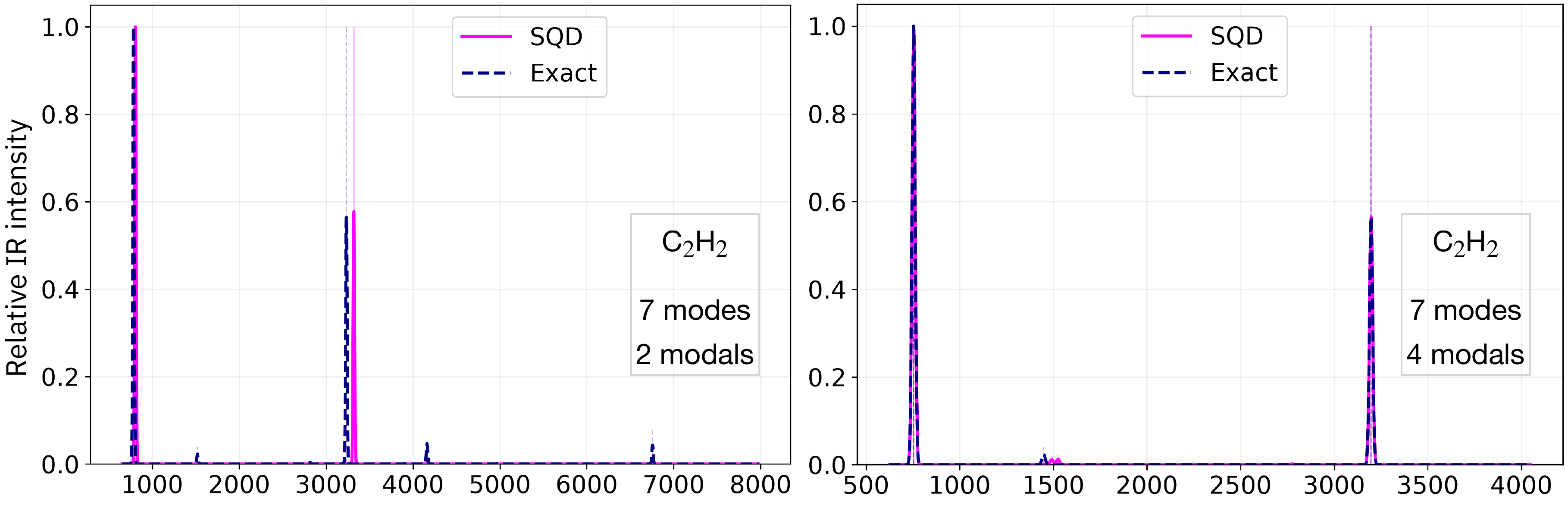}
    \includegraphics[width=1\linewidth]{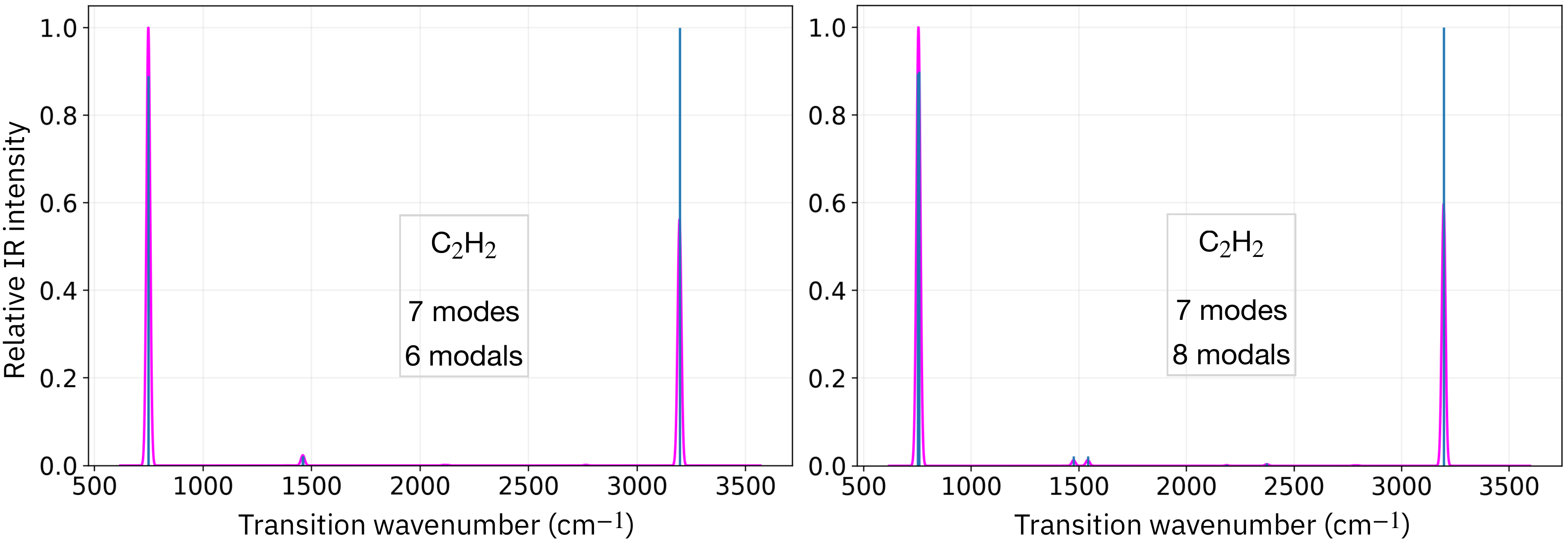}
    \caption{Relative IR spectra are presented based on the 0 $\rightarrow$ $m$ transition dipole moments for C$_2$H$_2$, with $m$ being up to 99 (ground state and 99 excited states), using the eigen-states obtained from  SQD procedure applied on samples measured from the vib-LUCJ circuits. In the top row, we compare the SQD result with the result obtained from exact diagonalization of the corresponding Hamiltonian.}
    \label{DipoleC2H2_lucj}
\end{figure*}

%%%%%%%%%%%%%%%%%%%%%%%%%%%%%%%%%%%%%%%%%%%%%%%%%%%%%%%%%%%%%%%%%%%%%%%%%
\begin{figure*}[h]
    \centering
    \includegraphics[width=1\linewidth]{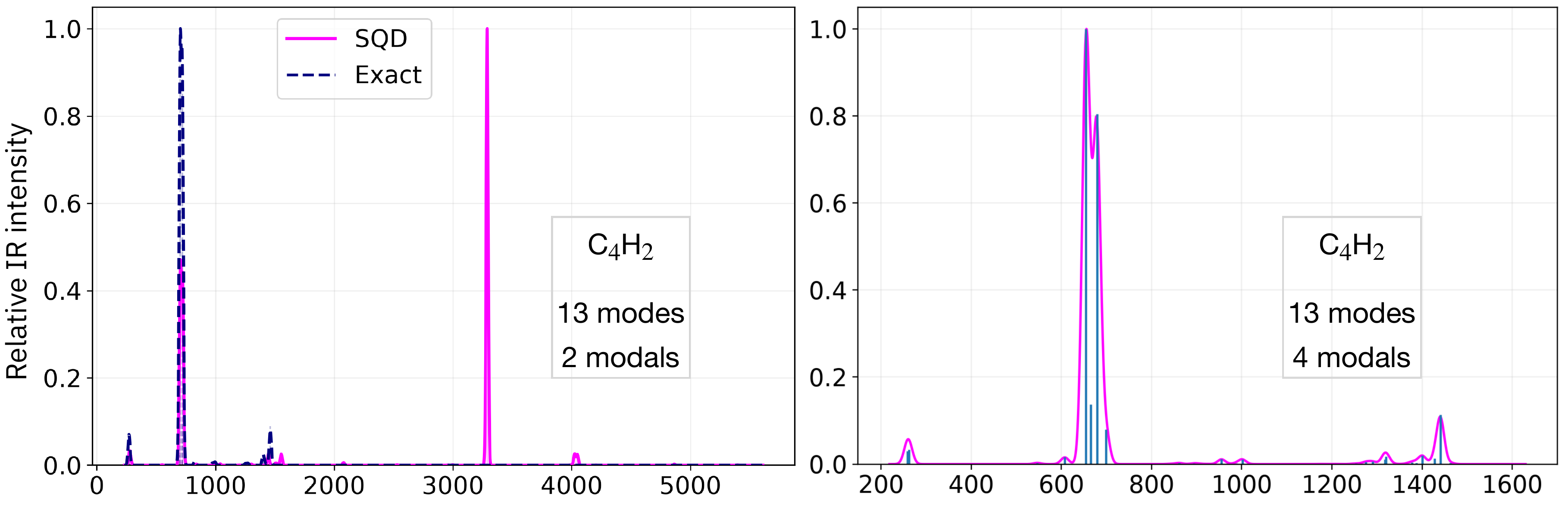}
    \includegraphics[width=1\linewidth]{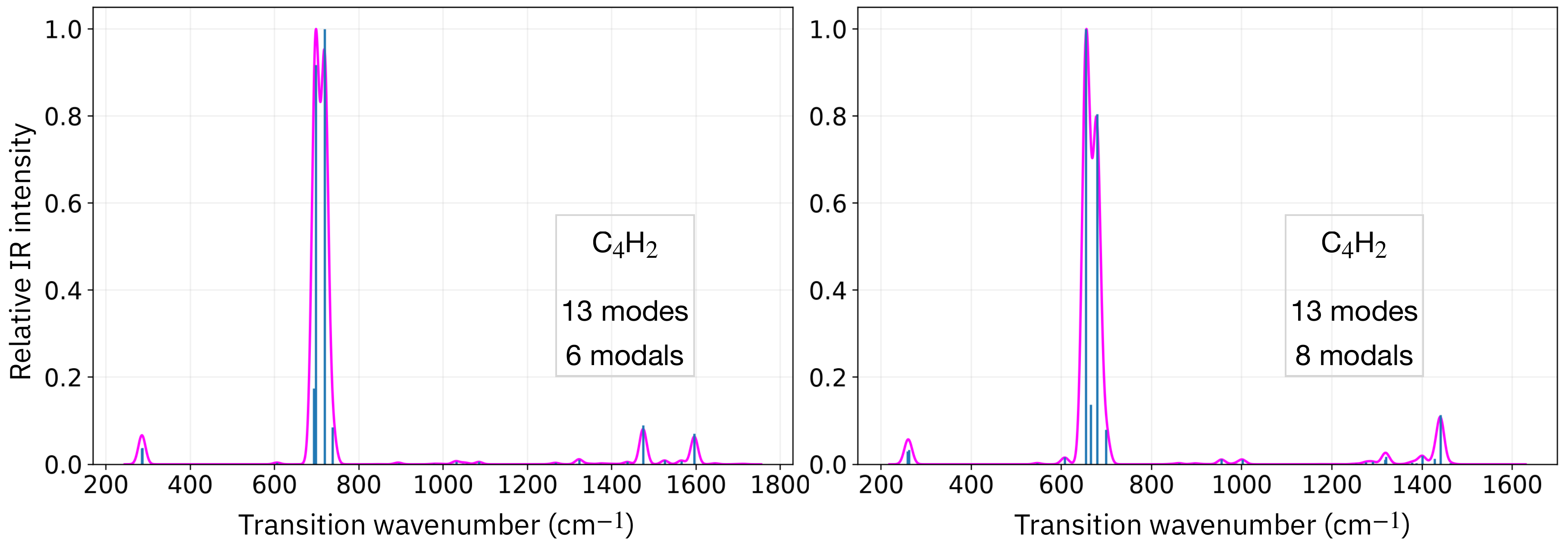}
    \caption{Relative IR spectra are presented based on the 0 $\rightarrow$ $m$ transition dipole moments for C$_4$H$_2$,  with $m$ being up to 99 (ground state and 99 excited states), using the eigen-states obtained from  SQD procedure applied on samples measured from the vib-LUCJ circuits. In the top-left panel, we compare the SQD result with the result obtained from exact diagonalization of the corresponding 2 modal Hamiltonian.}
    \label{DipoleC4H2_lucj}
\end{figure*}

%%%%%%%%%%%%%%%%%%%%%%%%%%%%%%%%%%%%%%%%%%%%%%%%%%%%%%%%%%%%%%%%%%%%%%%%%

\begin{figure*}[h]
    \centering
    \includegraphics[trim=0cm 6.5cm 0.5cm 6cm, clip, width=1\linewidth]{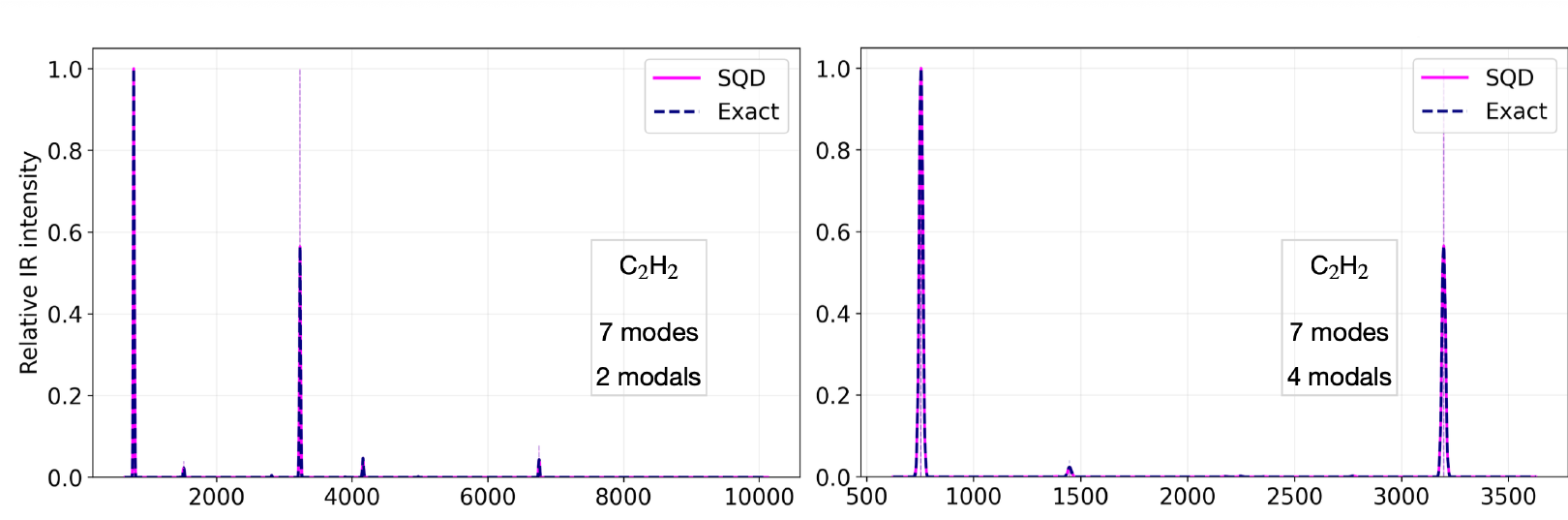}
    \includegraphics[trim=0cm 6cm 0.45cm 6.5cm, clip, width=1\linewidth]{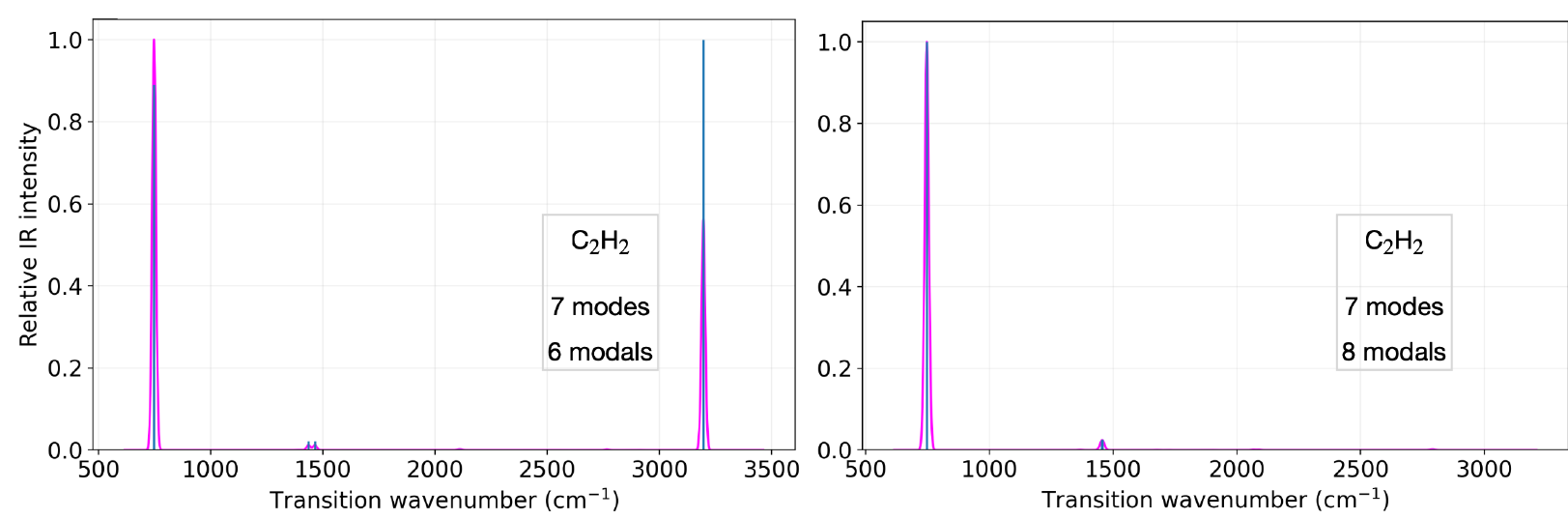}
    \caption{Relative IR spectra are presented based on the 0 $\rightarrow$ $m$ transition dipole moments for C$_2$H$_2$,  with $m$ being up to 49 (ground state and 49 excited states), using the eigen-states obtained from SQD procedure applied on samples measured from the vib-SqDRIFT circuits. Results obtained for the lowest ground state energy were used. In the top row, we compare the SQD result with the result obtained from exact diagonalization of the corresponding modal Hamiltonians.}
    \label{Dipole_C2H2_SqDrift}
\end{figure*}

%%%%%%%%%%%%%%%%%%%%%%%%%%%%%%%%%%%%%%%%%%%%%%%%%%%%%%%%%%%%%%%%%%%%%%%%%
\begin{figure*}[h]
    \centering
    \includegraphics[trim=0cm 6.5cm 0.5cm 6cm, clip, width=1\linewidth]{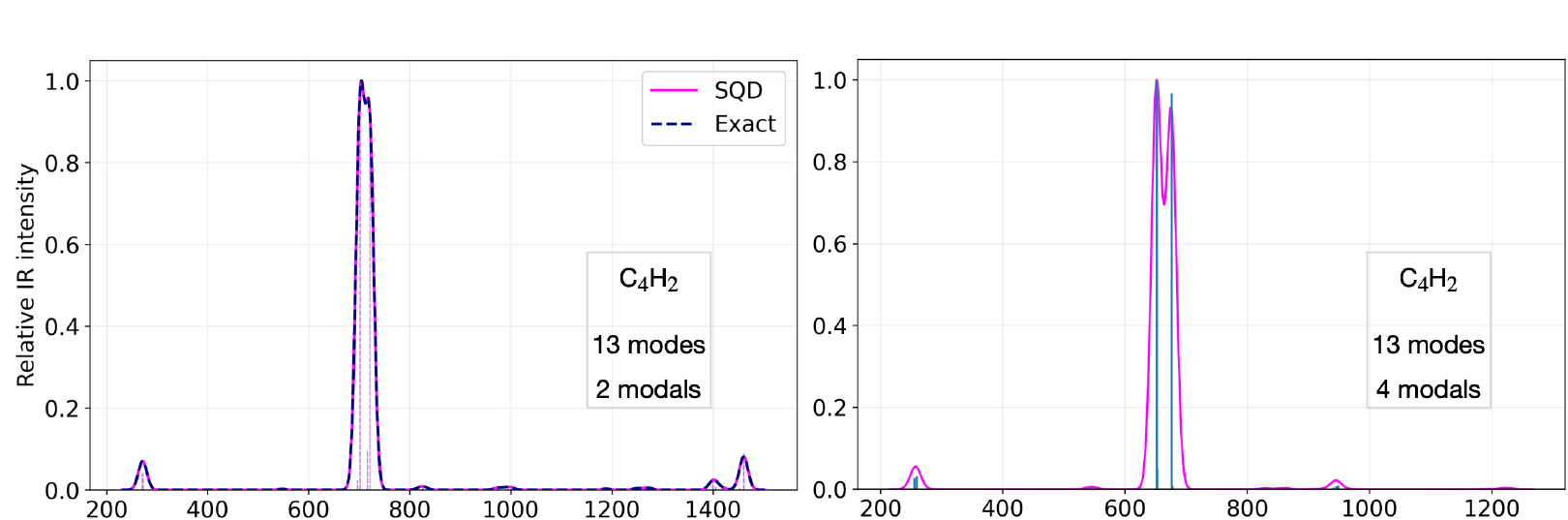}
    \includegraphics[trim=0cm 6cm 0.5cm 6.5cm, clip, width=1\linewidth]{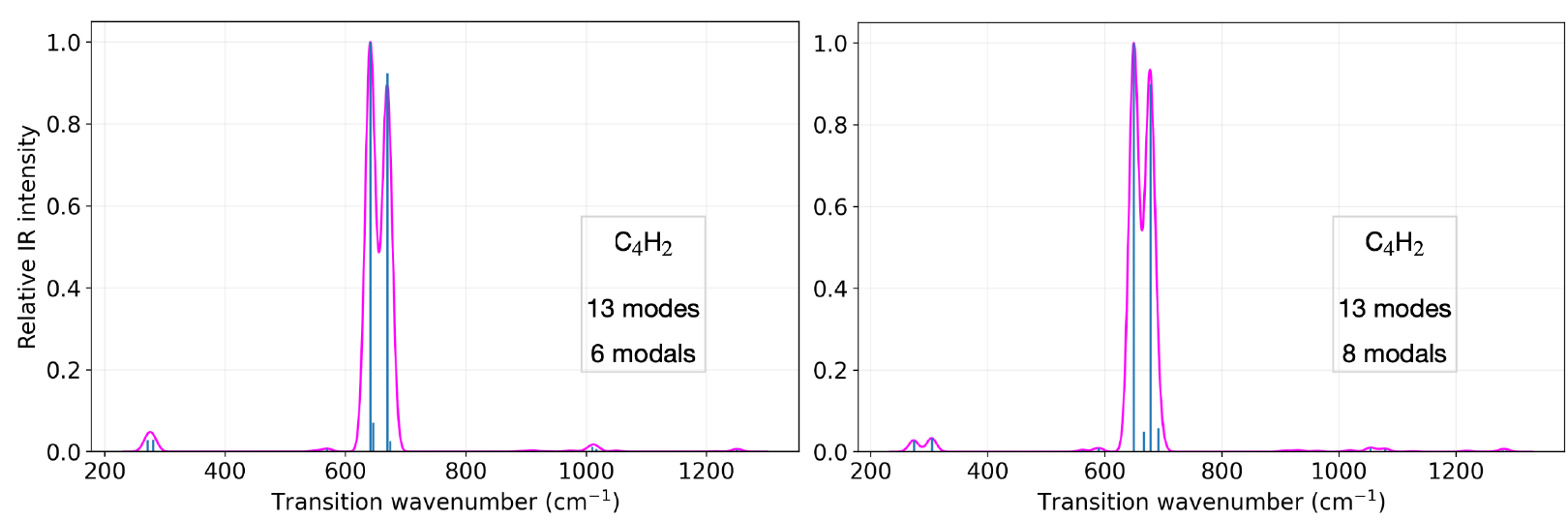}
    \caption{Relative IR spectra are presented based on the 0 $\rightarrow$ $m$ transition dipole moments for C$_4$H$_2$,  with $m$ being up to 49 (ground state and 49 excited states), using the eigen-states obtained from SQD procedure applied on samples measured from the vib-SqDRIFT circuits. Results obtained for the lowest ground state energy were used. In the top-left panel, we compare the SQD result with the result obtained from exact diagonalization of the corresponding 2 modal Hamiltonians.}
    \label{Dipole_C4H2_SqDrift}
\end{figure*}

%%%%%%%%%%%%%%%%%%%%%%%%%%%%%%%%%%%%%%%%%%%%%%%%%%%%%%%%%%%%%%%%%%%%%%%%%
%%%%%%%%%%%%%%%%%%%%%%%%%%%%%%%%%%%%%%%%%%%%%%%%%%%%%%%%%%%%%%%%%%%%%%%%%
\subsection{Infrared spectra and modal-basis convergence}
\label{dipoleTransition}
Having established the convergence of the low-energy eigenvalues, we next examine whether the recovered SQD wavefunctions are sufficiently accurate to reproduce infrared transition properties. This provides a more demanding benchmark since transition intensities depend on matrix elements between the ground and excited vibrational states rather than on a single eigenvalue. Figures~\ref{DipoleC2H2_lucj}-\ref{Dipole_C4H2_SqDrift} compare the resulting zero-temperature IR spectra as the number of retained modal basis functions is increased.

With the transition dipole between the vibrational ground state and an excited
state $m$ being \mbox{$\mu_\chi^{m0}=  \bra{\Psi_m}\mu_\chi\ket{\Psi_0} $,}
for $\chi = x, y, z $, the dipole strength is given by  ${\mathcal S}_{m0}= \sum_{\chi } \left|\mu_\chi^{m0}\right|^2 ,$ as we have discussed in Eq.~\ref{eq:dipoleStrength}.
At zero temperature, only the vibrational ground state is populated, and
the relative zero-temperature absorption intensity for a transition $0\rightarrow m$ can be written, up to an overall constant, as
\begin{align}
    \label{eq:relative_ir}
    I_{0\rightarrow{m}}^{\textrm{rel}}
    =
    \frac{\tilde{\nu}_{m0} \, {\mathcal S}_{m0}}
    {\max_g \left(\tilde{\nu}_{g0}\, {\mathcal S}_{g0}\right)}\, ,
\end{align}
following the usual electric-dipole expression for one-photon absorption
\cite{Baiardi2013_TDVibronicSpectroscopy,Carbonniere2020_AnharmonicIRRamanSolids}. 
Here $g$ runs over all the excited states and $\max_g $ finds the strongest frequency-weighted transition in that spectrum. 
$\tilde{\nu}$ is the transition frequency (wavenumber) between the vibrational ground state 0 and excited state $m$, expressed in $\mathrm{cm}^{-1}$.

Since the normalization in Eq.~\eqref{eq:relative_ir} is performed
independently for each modal truncation, relative intensities from different
truncations are not on a common absolute scale.  
Consequently, convergence across the 2-, 4-, 6-, and 8-modal calculations is assessed primarily from the raw transition-dipole components and from ${\mathcal S}_{m0}$, together with the configuration content of the corresponding SQD eigenstates. 
This distinction is important because transition properties need not converge
monotonically with the size of the vibrational basis.  Enlarging a modal basis
changes the variational representation of both the initial and final
wavefunctions and introduces additional overtone and combination
configurations.  
In VSCF/VCI-type descriptions, anharmonic eigenstates are linear combinations of product-modal configurations, so both state energies and transition intensities can change as the modal/configuration space is enlarged \cite{Erba2019_VSCFVCISolids,Carbonniere2020_AnharmonicIRRamanSolids}.  
Anharmonic coupling can also redistribute oscillator strength among nearby states and can give observable intensity to combination bands through intensity borrowing
\cite{Yagi2008_QDPTFermiResonance,Carbonniere2020_AnharmonicIRRamanSolids}.

For C$_2$H$_2$, the SQD spectra reproduce the dominant features of exactly diagonalized truncated Hamiltonians in the basis sizes for which a direct comparison is available, see Figures~\ref{DipoleC2H2_lucj} and \ref{Dipole_C2H2_SqDrift}. The agreement provides a more stringent validation than the ground-state energy alone because both the transition frequencies and transition matrix elements depend on the reconstructed excited-state wavefunctions. As the number of modals is increased from two to eight, the dominant low- and high-frequency spectral features remain identifiable, while their relative positions and intensities evolve as additional anharmonic configurations enter the vibrational basis.

Linear diacetylene C$_4$H$_2$  provides a sensitive example of the dependence of the calculated infrared spectrum on the size of the local vibrational basis. As the number of modals per mode is increased, both the positions and the relative intensities of individual transitions change, and additional closely spaced spectral features become apparent. The evolution of the spectrum is therefore not described by a simple one-to-one correspondence between individual peaks at different modal truncations. Instead, enlargement of the modal basis changes the representation of the vibrational states and the distribution of transition strength among nearby transitions.
For the two-modal calculation, where exact diagonalization of the corresponding truncated Hamiltonian is feasible, the SQD spectrum can be compared directly with the exact result, providing a control on the transition properties obtained from the selected subspace. As the modal basis is subsequently enlarged, the detailed spectral structure continues to evolve even though the low-energy eigenvalues are comparatively stable. This illustrates that convergence of vibrational energies alone does not necessarily imply convergence of spectroscopic observables.
An additional consideration is that the spectra for different modal truncations are normalized independently. Consequently, relative peak heights obtained from different calculations are not on a common absolute scale and should not be interpreted directly as changes in absolute transition strength. Assessment of spectroscopic convergence should therefore consider the unnormalized transition strengths $S_{m0}$, or equivalently the underlying transition-dipole moments, together with the evolution of the peak positions, rather than normalized intensities alone.
The ground-to-excited-state transition dipole moments for C$_2$H$_2$ and C$_4$H$_2$ are provided in Appendix~\ref{Appendix:TransitionDipoles}, Figures~\ref{DipoleMomentC2H2_lucj}-\ref{DipoleMomentC4H2_SqDrift}. Their Cartesian components are evaluated from the SQD eigenstates and used to obtain the infrared intensities discussed here.

%%%%%%%%%%%%%%%%%%%%%%%%%%%%%%%%%%%%%%%%%%%%%%%%%%%%%%%%%%%%%%%%%%%%%%%%%
%%%%%%%%%%%%%%%%%%%%%%%%%%%%%%%%%%%%%%%%%%%%%%%%%%%%%%%%%%%%%%%%%%%%%%%%%
\section{Conclusions and Outlook}
\label{conclusion}
We have developed and tested an end-to-end sample-based quantum diagonalization framework for anharmonic vibrational Hamiltonians, combining quantum state preparation and sampling with configuration recovery, selected-subspace diagonalization, and evaluation of infrared transition properties. Two complementary sampling strategies, vib-LUCJ and vib-SqDRIFT, were applied to C$_2$H$_2$ and C$_4$H$_2$ while systematically increasing the number of basis functions retained for each vibrational mode. The resulting calculations extend to 56 qubits for C$_2$H$_2$ and 104 qubits for C$_4$H$_2$. The calculations show that low-energy vibrational states can be recovered from selected spaces that are substantially smaller than the corresponding full truncated product spaces. 
Both vib-LUCJ and vib-SqDRIFT identify compact subsets of the vibrational configuration space that retain the information required to reproduce low-energy states with good agreement to the classical reference calculations. Importantly, the sampling parameters used here, including the number of samples per batch, were not systematically optimized for maximizing the coverage of the relevant configuration space, leaving room for further improvement. 
Since VCC[2] and VCC[3] are themselves approximate reference methods, deviations from these energies should not be interpreted as absolute SQD errors. Comparison with systematically converged classical approaches such as VCI, where computationally feasible, would provide a more definitive assessment of the absolute accuracy of the SQD results.

The transition-property calculations provide an additional and more sensitive test of the reconstructed eigenspace. For systems and basis sizes where direct diagonalization of the corresponding truncated Hamiltonian remains feasible, the SQD spectra reproduce the dominant transition positions and intensity patterns, supporting the consistency of the recovered excited states and subsequent dipole-moment evaluation. At the same time, the C$_4$H$_2$ results show that increasing the modal basis can substantially alter individual peak positions, intensities, and excited-state composition even when the low-energy eigenvalues are comparatively stable. The state-resolved analysis further indicates that nearby low-frequency transitions originate from distinct modal families with markedly different strengths, while the summed unnormalized transition strength of the dominant family remains comparatively stable as the basis is enlarged. This demonstrates that convergence of vibrational energies alone does not necessarily imply convergence of spectroscopic observables.

Taken together, these results support sample-based subspace diagonalization as a promising approach for treating anharmonic vibrational problems without explicitly constructing the complete vibrational product basis. Future work should focus on systematic convergence with sampling resources, comparison against controlled classical selection strategies and higher-level vibrational benchmarks, and extension to larger molecular systems in which the full modal space is no longer tractable.

%%%%%%%%%%%%%%%%%%%%%%%%%%%%%%%%%%%%%%%%%%%%%%%%%%%%%%%%%%%%%%%%%%%%%%%%%
%%%%%%%%%%%%%%%%%%%%%%%%%%%%%%%%%%%%%%%%%%%%%%%%%%%%%%%%%%%%%%%%%%%%%%%%%
\section*{acknowledgement}

We thank our colleagues Mario Motta, Benjamin Jaderberg, Bob Parney and Alexandre Pfeifer for insightful discussions, as well as Matthew Stypulkoski for support with the IBM Quantum Hardware access. We are also grateful to Kunal Kumar and Yuri Kobayashi for sharing perspectives from their concurrent research on vibrational SQD methods. A related manuscript has been published in parallel.

%%%%%%%%%%%%%%%%%%%%%%%%%%%%%%%%%%%%%%%%%%%%%%%%%%%%%%%%%%%%%%%%%%%%%%%%%
%%%%%%%%%%%%%%%%%%%%%%%%%%%%%%%%%%%%%%%%%%%%%%%%%%%%%%%%%%%%%%%%%%%%%%%%%

\bibliographystyle{achemso}
\bibliography{bibliography.bib}

%%%%%%%%%%%%%%%%%%%%%%%%%%%%%%%%%%%%%%%%%%%%%%%%%%%%%%%%%%%%%%%%%%%%%%%%%
%%%%%%%%%%%%%%%%%%%%%%%%%%%%%%%%%%%%%%%%%%%%%%%%%%%%%%%%%%%%%%%%%%%%%%%%%

\newpage
\appendix 

\newpage

\section{Transition dipole moments}
\label{Appendix:TransitionDipoles}
The Cartesian components of the ground-to-excited-state transition dipole moments for C$_2$H$_2$ and C$_4$H$_2$ are provided in this Appendix. They are evaluated from the SQD eigenstates using the corresponding dipole-moment operators and are used to compute the infrared intensities presented in the main text.

%%%%%%%%%%%%%%%%%%%%%%%%%%%%%%%%%%%%%%%%%%%%%%%%%%%%%%%%%%%%%%%%%%%%%%%%%
%%%%%%%%%%%%%%%%%%%%%%%%%%%%%%%%%%%%%%%%%%%%%%%%%%%%%%%%%%%%%%%%%%%%%%%%%
\begin{figure*}[h]
    \centering
    \includegraphics[width=1\linewidth]{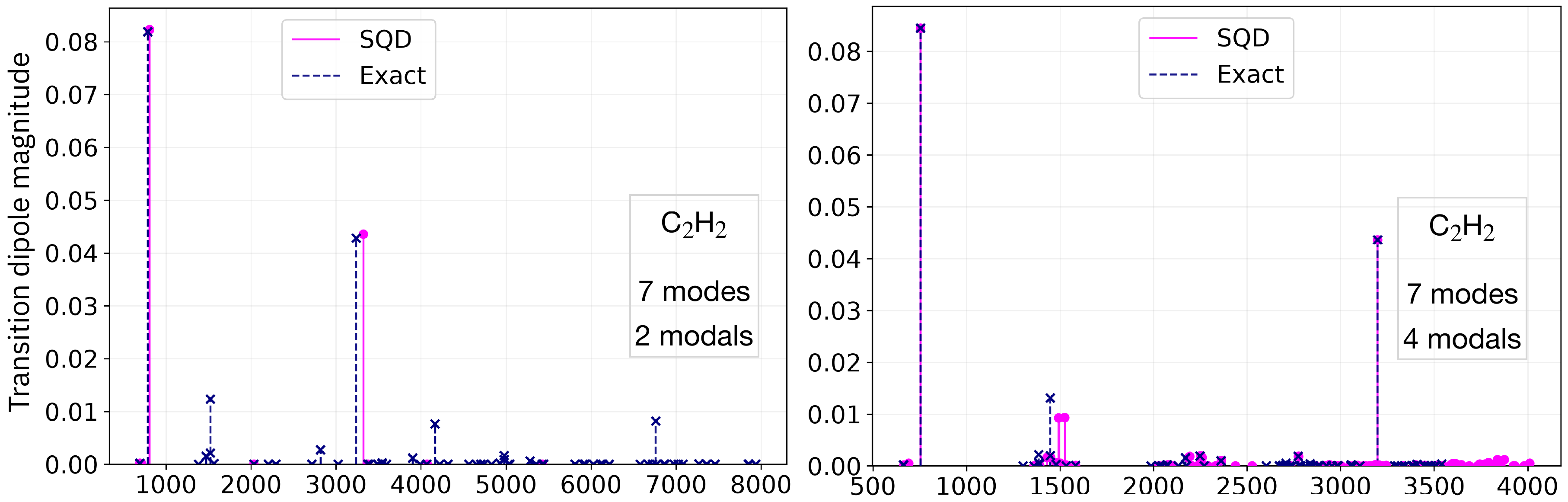}
    \includegraphics[width=1\linewidth]{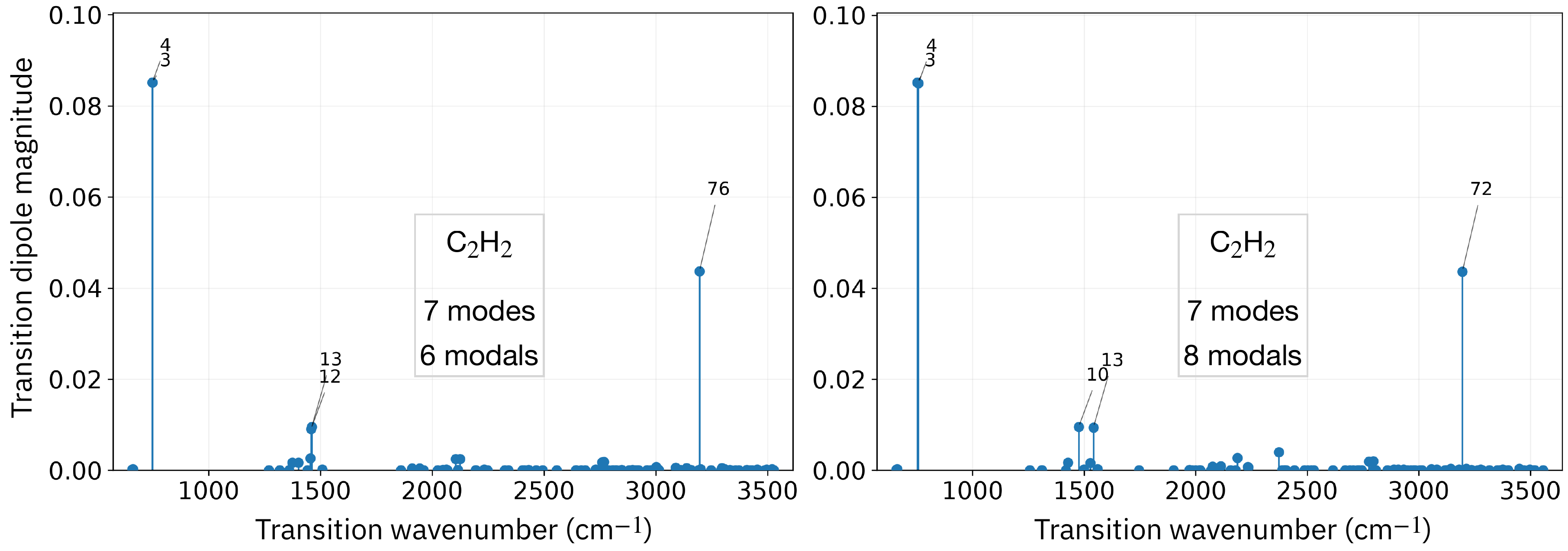}
    \caption{Transition dipole moments for C$_2$H$_2$ for 0 $\rightarrow$ $m$ transitions with $m$ being up to 99 (ground state and 99 excited states), using the eigen-states obtained from SQD procedure applied on samples measured from the vib-LUCJ circuits. In the top row, we compare the SQD result with the result obtained from exact diagonalization of the corresponding Hamiltonian.}
    \label{DipoleMomentC2H2_lucj}
\end{figure*}

%%%%%%%%%%%%%%%%%%%%%%%%%%%%%%%%%%%%%%%%%%%%%%%%%%%%%%%%%%%%%%%%%%%%%%%%%
\begin{figure*}[t]
    \centering
    \includegraphics[width=1\linewidth]{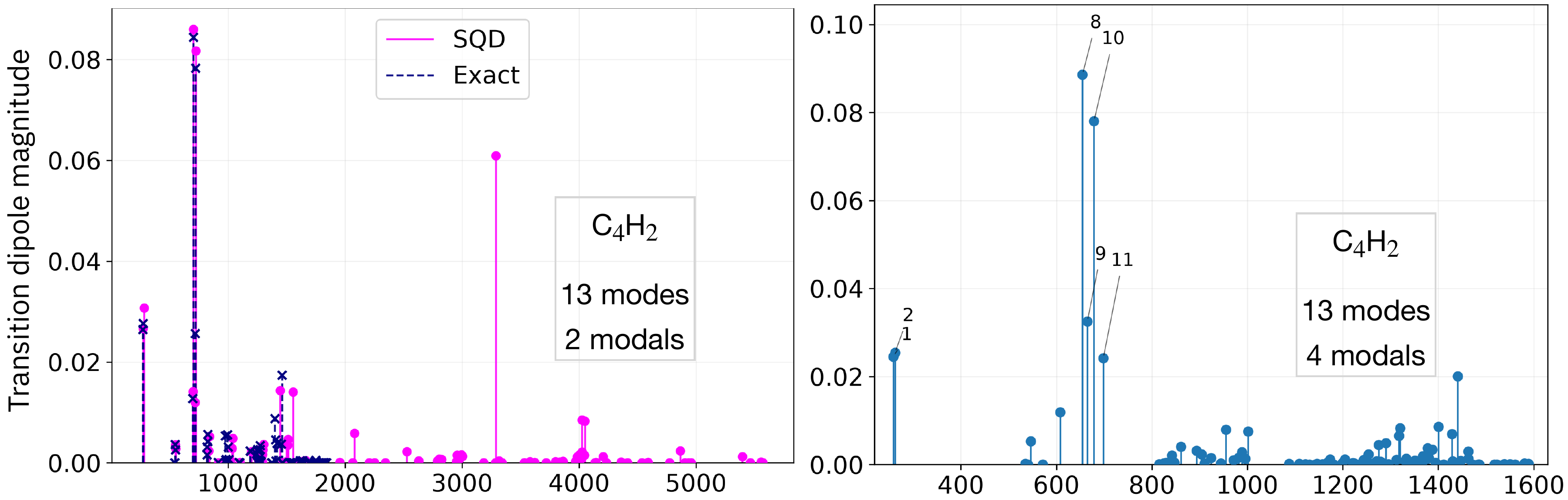}
    \includegraphics[width=1\linewidth]{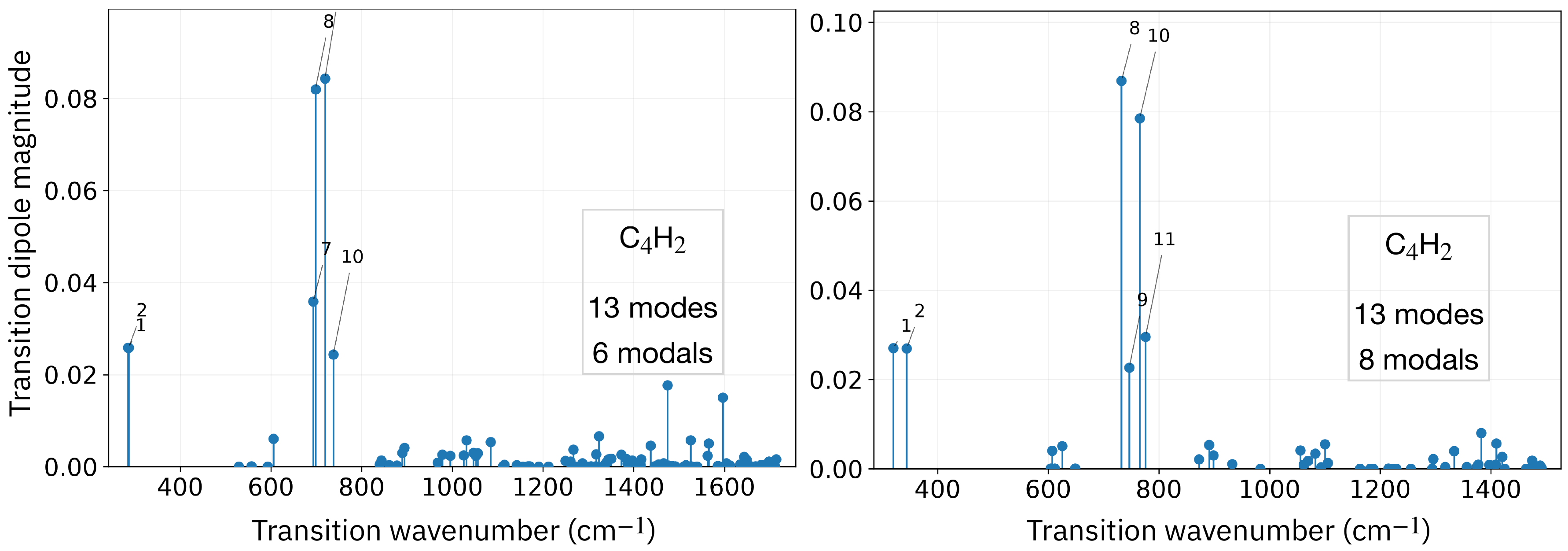}
    \caption{Transition dipole moments for C$_4$H$_2$ for 0 $\rightarrow$ $m$ transitions with $m$ being up to 99 (ground state and 99 excited states), using the eigen-states obtained from SQD procedure applied on samples measured from the vib-LUCJ circuits. In the top-left panel, we compare the SQD result with the result obtained from exact diagonalization of the corresponding 2 modal Hamiltonian.}
    \label{DipoleMomentC4H2_lucj}
\end{figure*}

%%%%%%%%%%%%%%%%%%%%%%%%%%%%%%%%%%%%%%%%%%%%%%%%%%%%%%%%%%%%%%%%%%%%%%%%%
\begin{figure*}[h]
    \centering
    \includegraphics[trim=0cm 5.6cm 0.4cm 5cm, clip, width=1\linewidth]{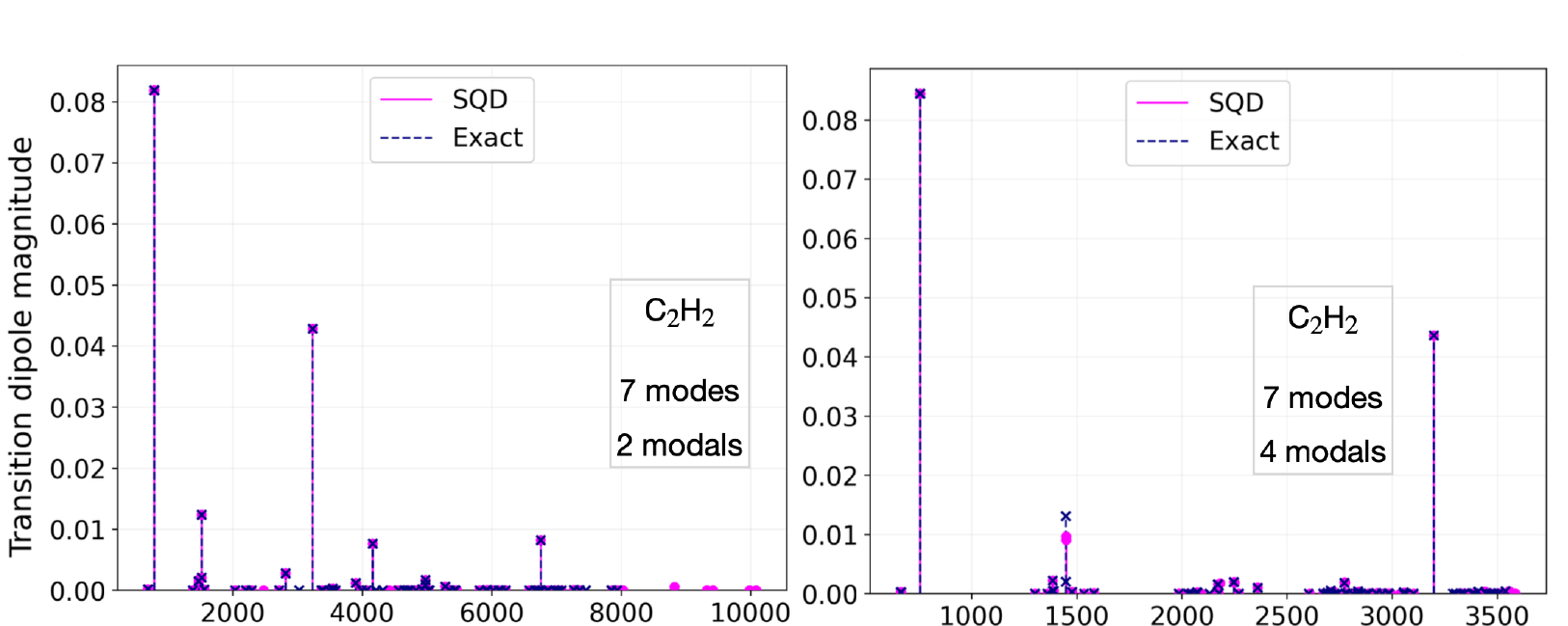}
    \includegraphics[trim=0cm 5cm 0.5cm 5.5cm, clip, width=1.005\linewidth]{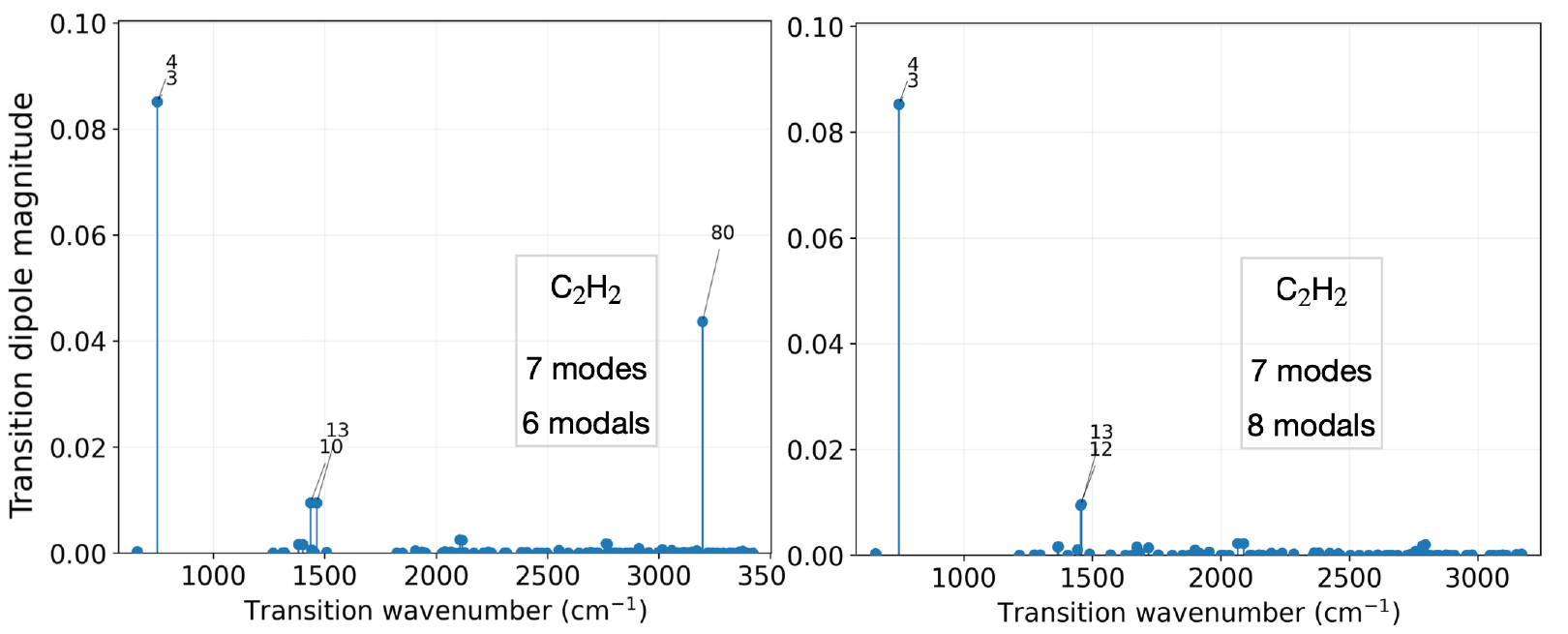}
    \caption{Transition dipole moments for C$_2$H$_2$ for 0 $\rightarrow$ $m$ transitions with $m$ being up to 99 (ground state and 99 excited states), using the eigen-states obtained from SQD procedure applied on samples measured from the vib-SqDRIFT circuits. In the top row, we compare the SQD result with the result obtained from exact diagonalization of the corresponding Hamiltonian.}
    \label{DipoleMomentC2H2_SqDrift}
\end{figure*}

%%%%%%%%%%%%%%%%%%%%%%%%%%%%%%%%%%%%%%%%%%%%%%%%%%%%%%%%%%%%%%%%%%%%%%%%%
\begin{figure*}[t]
    \centering
    \includegraphics[trim=0cm 5.6cm 0.4cm 5cm, clip, width=1\linewidth]{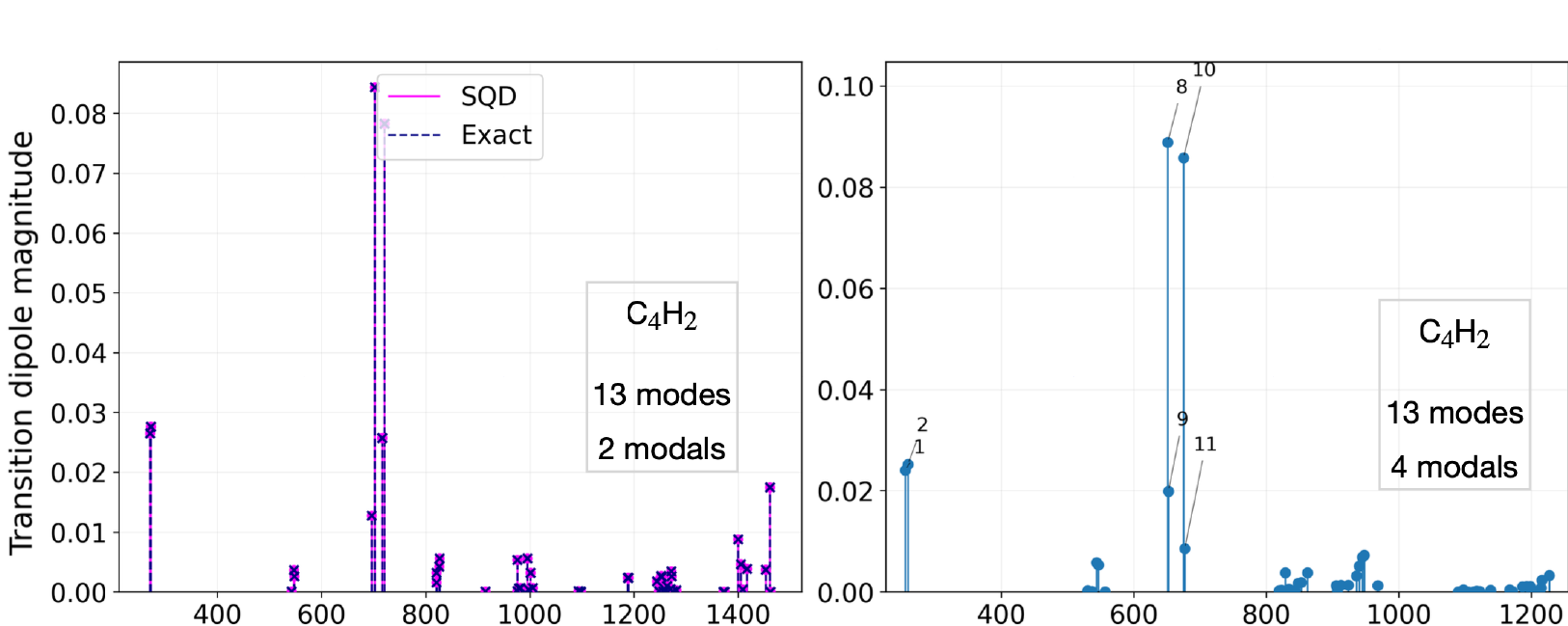}
    \includegraphics[trim=0cm 5cm 0.5cm 5.5cm, clip, width=1\linewidth]{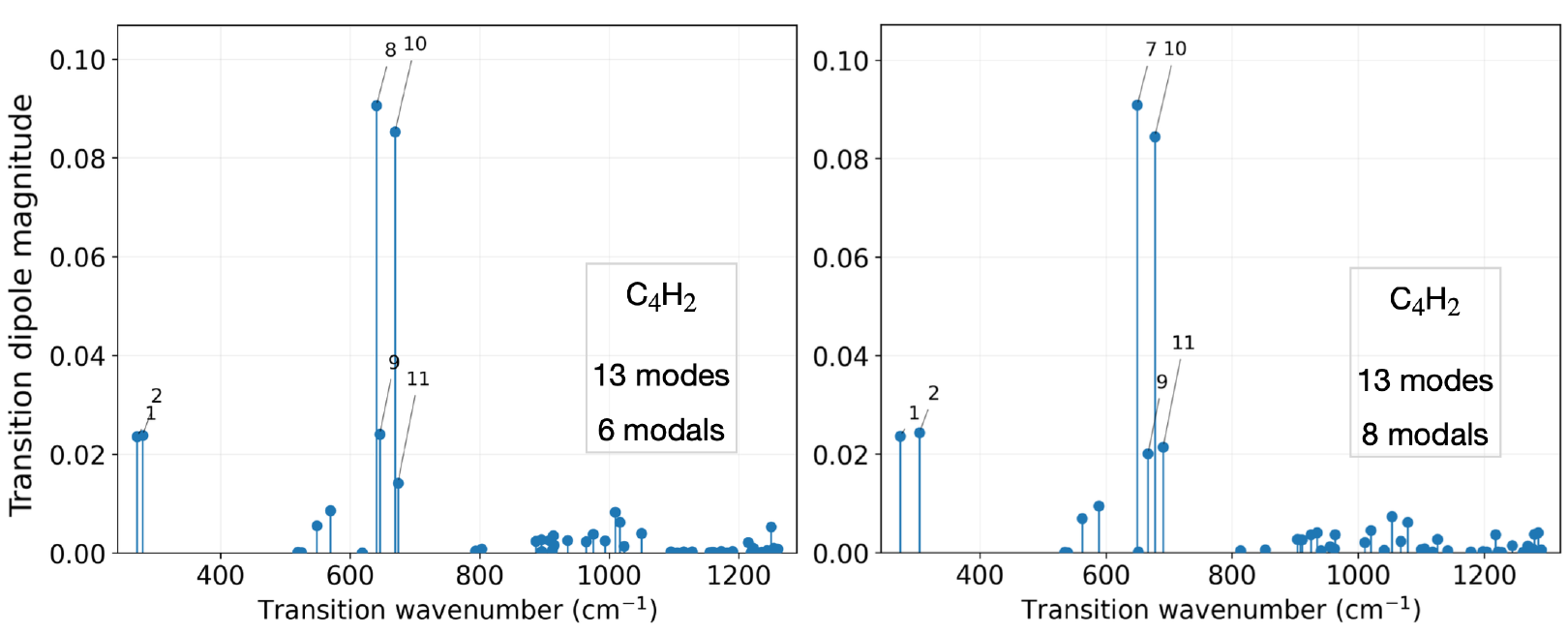}
    \caption{Transition dipole moments for C$_4$H$_2$ for 0 $\rightarrow$ $m$ transitions with $m$ being up to 99 (ground state and 99 excited states), using the eigen-states obtained from SQD procedure applied on samples measured from the vib-SqDRIFT circuits. In the top-left panel, we compare the SQD result with the result obtained from exact diagonalization of the corresponding 2 modal Hamiltonian.}
    \label{DipoleMomentC4H2_SqDrift}
\end{figure*}

%%%%%%%%%%%%%%%%%%%%%%%%%%%%%%%%%%%%%%%%%%%%%%%%%%%%%%%%%%%%%%%%%%%%%%%%%
%%%%%%%%%%%%%%%%%%%%%%%%%%%%%%%%%%%%%%%%%%%%%%%%%%%%%%%%%%%%%%%%%%%%%%%%%

\newpage

\section{Circuit Resources}
\label{Circuit_Resources}
Circuit resources required for running vib-LUCJ and  vib-SqDRIFT circuits after transpilation to the \texttt{ibm\_aachen} and \texttt{ibm\_phoenix} devices, respectively, are listed below in Table \ref{Tab:lucj_circ_res} and \ref{Tab:SqDRIFT_circ_res}.

%%%%%%%%%%%%%%%%%%%%%%%%%%%%%%%%%%%%%%%%%%%%%%%%%%%%%%%%%%%%%%%%%%%%%%%%%
%%%%%%%%%%%%%%%%%%%%%%%%%%%%%%%%%%%%%%%%%%%%%%%%%%%%%%%%%%%%%%%%%%%%%%%%%
\begin{table}[h]
\centering
\caption{
\label{Tab:lucj_circ_res}
Quantum circuit resources required after transpiling LUCJ circuits to the \texttt{ibm\_aachen} quantum processor for the systems $\text{C}_{2}\text{H}_{2}$ and $\text{C}_{4}\text{H}_{2}$.}
\begin{tabular}{|c|c|c|c|c|c|c|} 
\hline
Molecule & modes & modals &  qubits & Single qubit gates & Two qubit gates & Circuit Depth  \\ \hline\hline
C$_2$H$_2$ & 7 & 2 & 14  & 60 & 7 &  16 \\ \hline 
C$_2$H$_2$ & 7 & 4 &  28  & 2743  & 514 & 537 \\ \hline 
C$_2$H$_2$ & 7 & 6 &  42  &  6665 & 1411 &  1086 \\ \hline
C$_2$H$_2$ & 7  & 8 &  56  & 7048  & 1697 &  1603 \\ \hline
C$_4$H$_2$ & 13 & 2 &  26  &  162 & 24 & 16 \\ \hline 
C$_4$H$_2$ & 13 & 4 &  52  & 11386  & 2361 & 1446 \\ \hline 
C$_4$H$_2$ & 13 & 6 &  78  & 22769  & 5181 & 1838 \\ \hline
C$_4$H$_2$ & 13 & 8 & 104  &  31336  & 7390 &  4172 \\ \hline
\end{tabular}
\end{table}

%%%%%%%%%%%%%%%%%%%%%%%%%%%%%%%%%%%%%%%%%%%%%%%%%%%%%%%%%%%%%%%%%%%%%%%%%
%%%%%%%%%%%%%%%%%%%%%%%%%%%%%%%%%%%%%%%%%%%%%%%%%%%%%%%%%%%%%%%%%%%%%%%%%

\begin{table}[h]
\centering
\caption{
\label{Tab:SqDRIFT_circ_res}
Quantum circuit resources required after transpiling SqDRIFT circuits to the \texttt{ibm\_phoenix} quantum processor for the systems $\text{C}_{2}\text{H}_{2}$ and $\text{C}_{4}\text{H}_{2}$ averaged over 100 randomizations and 4 Krylov time steps excluding the VSCF state at t=0. 
%The circuit parameters that give the lowest energy among the excitations are highlighted.
}
\begin{tabular}{|c|c|c|c|c|c|c|c|} 
    \hline
    Molecule & modes & modals & $N_{exc}$ & qubits & Single qubit gates & Two qubit gates & Circuit Depth  \\ \hline\hline
C$_2$H$_2$ & 7 & 2 & 25  & 14  & 345 & 98 &  211 \\  \hline 
C$_2$H$_2$ & 7 & 2 & 50  & 14  & 648 & 185 & 400 \\  \hline 
C$_2$H$_2$ & 7 & 2 & 100 & 14  & 1243  & 355 & 775 \\  \hline 
C$_2$H$_2$ & 7 & 2 & 250 & 14  & 2986  & 844 & 1873 \\  \hline 
C$_2$H$_2$ & 7 & 4 & 25  & 28  &  1734 & 545 & 960 \\ \hline 
C$_2$H$_2$ & 7 & 4 & 50  & 28  & 3430  & 1075 & 1914\\ \hline 
C$_2$H$_2$ & 7 & 4 & 100 & 28  & 6741 & 2111 & 3816 \\ \hline 
C$_2$H$_2$ & 7 & 4 & 250 & 28  & 16543  & 5175 & 9598\\ \hline 
C$_2$H$_2$ & 7 & 6 & 25  & 42  & 2313  & 729 &  1181 \\ \hline
C$_2$H$_2$ & 7 & 6 & 50  & 42  & 4543  & 1431 & 2329  \\ \hline
C$_2$H$_2$ & 7 & 6 & 100 & 42  & 9054  & 2861 & 4655 \\ \hline
C$_2$H$_2$ & 7 & 6 & 250 & 42  & 22720  & 7169 &  11712 \\ \hline
C$_2$H$_2$ & 7  & 8 & 25 & 56  & 2632  & 832 & 1214  \\ \hline
C$_2$H$_2$ & 7  & 8 & 50 & 56  & 5196  & 1642 & 2425  \\ \hline
C$_2$H$_2$ & 7  & 8 & 100& 56  & 10322  & 3274 & 4832 \\ \hline
C$_2$H$_2$ & 7  & 8 & 250& 56  & 25910  & 8222 &  12186 \\ \hline
C$_4$H$_2$ & 13 & 2 & 25 & 26  & 530  & 154 & 249\\ \hline 
C$_4$H$_2$ & 13 & 2 & 50 & 26  & 1001  & 290 &  468 \\ \hline 
C$_4$H$_2$ & 13 & 2 & 100& 26  & 1941 & 564 & 932 \\ \hline 
C$_4$H$_2$ & 13 & 2 & 250& 26  & 4800  & 1380 &  2343 \\ \hline 
C$_4$H$_2$ & 13 & 4 & 25 & 52  & 2071  & 644 & 932 \\ \hline 
C$_4$H$_2$ & 13 & 4 & 50 & 52  & 4099  & 1278 & 1883 \\ \hline 
C$_4$H$_2$ & 13 & 4 & 100& 52  & 8160  & 2553 & 3777\\ \hline 
C$_4$H$_2$ & 13 & 4 & 250& 52  & 20515  & 6417 & 9491 \\ \hline 
C$_4$H$_2$ & 13 & 6 & 25 & 78  & 2807 & 881 & 1069 \\ \hline
C$_4$H$_2$ & 13 & 6 & 50 & 78  & 5554  & 1745  &  2185 \\ \hline
C$_4$H$_2$ & 13 & 6 & 100& 78  & 11029  & 3479 &  4382 \\ \hline
C$_4$H$_2$ & 13 & 6 & 250& 78  & 27643  & 8736 & 11033 \\ \hline
C$_4$H$_2$ & 13 & 8 & 25 & 104  & 3334  & 1056 & 1096  \\ \hline
C$_4$H$_2$ & 13 & 8 & 50 & 104  & 6612 & 2088 &  2220 \\ \hline
C$_4$H$_2$ & 13 & 8 & 100& 104  & 13050  & 4127 & 4517 \\ \hline
C$_4$H$_2$ & 13 & 8 & 250& 104  & 32448 & 10290 &  11305 \\ \hline
  \end{tabular}
\end{table}

%%%%%%%%%%%%%%%%%%%%%%%%%%%%%%%%%%%%%%%%%%%%%%%%%%%%%%%%%%%%%%%%%%%%%%%%%
%%%%%%%%%%%%%%%%%%%%%%%%%%%%%%%%%%%%%%%%%%%%%%%%%%%%%%%%%%%%%%%%%%%%%%%%%
%\begin{table}[t]
%    \centering
%  \begin{tabular}{|c|c|c|c|c|c|c|} 
%    \hline
%    Molecule & modes & modals & qubits (unary) & ansatz & $E_{\rm SQD}$ & $E_{\rm SQD}$ - $E_{\rm Exact}$\\ \hline\hline
%C$_2$H$_4$ & 12 & 2 & 24 & LUCJ &  0.05022766 E$_h$ & 0.006030 cm$-1$   \\ \hline 
%C$_2$H$_4$ & 12 & 2 & 24 & QDRIFT &   & \\ \hline
%C$_2$H$_4$ & 12 & 4 & 48 & LUCJ &  0.05008458 E$_h$ &  \\ \hline  
%C$_2$H$_4$ & 12 & 4 & 48 & QDRIFT  &  & \\ \hline
%N$_2$H$_4$ & 12 & 2 & 24 & LUCJ &  0.05238878 E$_h$ & 0.000879 cm$-1$   \\ \hline
%N$_2$H$_4$ & 12 & 2 & 24 & QDRIFT &  & \\ \hline
%N$_2$H$_4$ & 12 & 4 & 48 & LUCJ & 0.05205609 E$_h$  &  \\ \hline
%N$_2$H$_4$ & 12 & 4 & 48 & QDRIFT &  &  \\ \hline
%  \end{tabular}
%\caption{Number of qubits for the unary encoding $N_q = Ld$, with $L$ being  number of the vibrational modes and $d$ number of the modals. Number of the vibrational modes is given by $L=3N-6$ for nonlinear molecules and $L=3N-5$ for the linear ones, with $N$ being the number of atoms. }
%\end{table}

\end{document}